\documentclass[aps, pra, twocolumn, superscriptaddress, amsmath,  tightenlines, longbibliography]{revtex4-2}

\usepackage{dcolumn}
\usepackage{graphicx}
\usepackage{epstopdf}
\usepackage{mathrsfs}
\usepackage{subfigure}
\usepackage{booktabs}
\usepackage{amsmath,amsfonts}
\usepackage{physics}
\usepackage{dsfont}
\usepackage{amstext}
\usepackage{amssymb}
\usepackage{amsbsy}
\usepackage{bbm}
\usepackage{amsthm}
\usepackage{graphicx}
\usepackage{xcolor}
\usepackage[colorlinks,urlcolor=blue,linkcolor=blue,citecolor=blue]{hyperref}
\usepackage{soul}

\usepackage{url}
\usepackage[colorlinks]{hyperref}
\hypersetup{%
	plainpages=true,
	breaklinks=true,       %not default in dvips mode, so we must specify
	hypertexnames=false,  %not ideal, but needed when pagenums duplicate (`i' vs. `1')
	pageanchor=true,
	colorlinks=true,
	linkcolor={blue},
	citecolor={blue},
	urlcolor={blue},
	anchorcolor={black}
}

 \makeatletter

\newcommand{\Rmnum}[1]{\expandafter\@slowromancap\romannumeral #1@}
\makeatother

\begin{document}
%\begin{CJK}{GBK}{song}
\title{Numerically-Exact Quantum-Simulation Approach for Two-Dimensional Spectroscopy of Open Quantum Systems}

\author{Yi-Xuan Yao}
\affiliation{School of Physics and Astronomy, Applied Optics Beijing Area Major Laboratory, Beijing Normal University, Beijing 100875, China}
\affiliation{Key Laboratory of Multiscale Spin Physics, Ministry of Education, Beijing Normal University, Beijing 100875, China}

\author{Hao-Yue Zhang}
\affiliation{School of Physics and Astronomy, Applied Optics Beijing Area Major Laboratory, Beijing Normal University, Beijing 100875, China}
\affiliation{Key Laboratory of Multiscale Spin Physics, Ministry of Education, Beijing Normal University, Beijing 100875, China}

\author{Cheng-Ge Liu}
\affiliation{School of Physics and Astronomy, Applied Optics Beijing Area Major Laboratory, Beijing Normal University, Beijing 100875, China}
\affiliation{Key Laboratory of Multiscale Spin Physics, Ministry of Education, Beijing Normal University, Beijing 100875, China}

\author{Rong-Hang Chen}
\affiliation{Beijing Computational Science Research Center, Beijing 100193, China}
\affiliation{School of Physics and Astronomy, Applied Optics Beijing Area Major Laboratory, Beijing Normal University, Beijing 100875, China}
\affiliation{Key Laboratory of Multiscale Spin Physics, Ministry of Education, Beijing Normal University, Beijing 100875, China}

\author{Qing Ai}
\email{aiqing@bnu.edu.cn}
\affiliation{School of Physics and Astronomy, Applied Optics Beijing Area Major Laboratory, Beijing Normal University, Beijing 100875, China}
\affiliation{Key Laboratory of Multiscale Spin Physics, Ministry of Education, Beijing Normal University, Beijing 100875, China}

\author{Franco Nori}
\email{fnori@riken.jp}
\affiliation{Theoretical Quantum Physics Laboratory, Cluster for Pioneering Research, RIKEN, Wakoshi, Saitama 351-0198, Japan}
\affiliation{Quantum Computing Center, RIKEN, Wakoshi, Saitama 351-0198, Japan}
\affiliation{Department of Physics, The University of Michigan, Ann Arbor, MI, 48109-1040, USA}

\date{\today}% It is always \today, today,

\begin{abstract}
Two-dimensional spectroscopy (2DS) is a powerful ultrafast technique for probing electronic and vibrational dynamics in complex microscopic systems. Extracting detailed information on system dynamics and system-bath interactions from 2DS experiments requires precise theoretical simulations for comparison, which motivates the development of numerically-exact and computationally-efficient simulation approaches. Here, we propose a quantum-simulation approach for 2DS based on the bath-engineering technique (BET), which has been successfully employed in quantum simulations of open quantum dynamics. To demonstrate our approach, we first simulate the 2DS of a driven four-level system in chiral enantiodetection, where we also assess the applicability of the center-line slope (CLS) method for extracting time correlation functions (TCFs) from the 2DS. We further apply our approach to the 2DS of ${\rm Rh(CO)_2C_5H_7O_2}$ (RDC) dissolved in chloroform, where the results reproduce the main spectral patterns observed in experiments. Our work provides a numerically-exact and efficient framework for simulating 2DS, and can offer additional insight into the dynamics of open quantum systems. %To demonstrate our approach, we simulate 2DSs of a driven four-level system for chiral enantiodetection and of ${\rm Rh(CO)_2C_5H_7O_2}$ dissolved in chloroform. The results for the latter are consistent with previous experimental results. Additionally, we validate the center-line slope (CLS) theory by applying it to the simulated 2DS, and observe that the extracted time correlation function (TCF) fits well with the preset one in the absorptive 2DS.  Our work thus provides an exact and efficient framework for simulating 2DSs, facilitating a deeper analysis and understanding of the information they contain. %Furthermore, we extract the time correlation functions (TCFs) from the simulated 2DS using the center-line slope (CLS) method. The extracted TCFs closely match the expected theoretical profiles, validating our approach. 
\end{abstract}

\maketitle

%\tableofcontents

\section{Introduction}
{Two-dimensional spectroscopy (2DS) is a powerful technique for probing the ultrafast processes in complex systems, with applications spanning physical \cite{Li2006PRL,Yang2008PRL,Lemmer2015PRL,Boss2016PRL}, chemical \cite{Kolano2006NA,Eaves2005pnas,Loparo2006JCP,Park2007pnas,Demirdoven2002PRL,Khalil2003JPCA,Fulmer2005PRL,Ruetzel2013PRL}, and biological \cite{Lambert2013NP,Cao2020SA,collini2010NA,Engel07Nature,BrixnerNA2005,ReadBJ2008,DostalJACS2012,Abramavicius2009CR} domains. By correlating excitation and detection frequencies in a time-resolved manner, 2DS provides insights into processes such as solute-solvent interactions, energy/charge transfer, and excitonic interactions that go beyond what linear spectroscopies can reveal \cite{Aue1976,Ernst1989,Mukamel1999,Hamm2011,Cho2009,Mukamel00ARPC,JonasARPC2003,ChoCR2008}. To interpret such experimental results, theoretical simulations serve as a bridge between observed spectral signatures and the underlying microscopic dynamics \cite{Tanimura2009ACR,Schlau-Cohen2012NC,Zhuang2006JPCB}.

One of the main challenges in simulating 2DS lies in the treatment of environmental effects \cite{Ishizaki2009JCP,Tanimura2020JCP}. Quantum systems are generally coupled to structured baths with diverse time-correlation functions (TCFs), which significantly influence the peak shapes and dynamical features in 2DS. Conventional perturbative approaches, such as the Redfield and F{\"o}rster theories, provide valuable insights within their respective validity regimes, but their underlying approximations limit their applicability to more complex systems \cite{Yang2002CP,Zigmantas2006PNAS,Lewis2013JPCA}. The hierarchical equations of motion (HEOM) provide a numerically-exact and widely-adopted framework for describing system-bath interactions \cite{Tanimura1989JPSJ,Tanimura2006JPSJ,Ishizaki2009JCP,Tanimura2020JCP,Ishizaki2005JPSJ,Tanimura2014JCP}, and have greatly advanced our understanding of open quantum dynamics. Nevertheless, the computational cost of the HEOM increases rapidly with the system dimensionality and the complexity of the spectral density, motivating the exploration of more-efficient numerical approaches in these conditions.
%Recently, the non-Hermitian approach has been developed as a more-economical strategy \cite{Lemmer2015PRL,Zhang2025JCTC}, offering a convenient way to capture spectral features and reproduce characteristic peak shapes with reduced effort, yet it remain less general. These limitations highlight the need for simulation frameworks that are both reliable and computationally efficient.

To this end, we propose a quantum-simulation approach for the 2DS of open quantum systems based on the bath-engineering technique (BET). Quantum-simulation approaches utilizing the BET have been recently proposed for open quantum systems with general spectral densities \cite{Soare2014PRA,zhang2021FOP} and experimentally realized in ion traps \cite{Soare2014NP,Soare2014PRA} and nuclear magnetic resonance \cite{Wang2018NPJQI,Chen2022NPJQI,Long2022PRL}. A key advantage of the BET is its flexibility with respect to the spectral-density profile, which facilitates simulations in highly-structured environments. Moreover, the computational cost in quantum-simulation paradigms is not expected to exhibit exponential scaling with respect to the system dimension \cite{Buluta2009S,Georgescu2014RMP}. These considerations suggest that the BET-based quantum simulation offers an efficient route for larger systems coupled to more complicated environments. 
As a further application, we use the BET-simulated 2DS to assess the center-line slope (CLS) theory for extracting the TCF of the bath from the 2DS \cite{Kwak2007JCP}.  By comparing the TCF extracted via the CLS from the BET-simulated 2DS with the preset TCF, we illustrate the practical applicability of the CLS theory. 

The article is structured as follows. We first introduce the BET in Sec.~\ref{sec:Method}. Next, we demonstrate the simulation of the 2DS in a four-level model relevant to enantiodetection of chiral molecules in Sec.~\ref{sec:Chiral}. The model consists of a cyclic three-level subsystem and a ground state. We then assess the applicability of the CLS theory for extracting the TCF in Sec.~\ref{sec:CLS}. Furthermore, we simulate the experimental 2DS of ${\rm Rh(CO)_2C_5H_7O_2}$ (RDC) dissolved in chloroform in Sec.~\ref{sec:RDC}, reproducing the main structural features observed in experiments \cite{Demirdoven2002PRL,Khalil2003JPCA}. 
Finally, we conclude our main findings in Sec.~\ref{sec:Conclusion}.
For the self-consistency, in Appendix~\ref{app:TCF}, we derive the TCF of the bath.
We further show the relation between the CLS and the TCF in Appendix~\ref{app:cls}. 
In order simulate the rephasing and non-rephasing 2DS by the BET, we provide the theoretical derivation in Appendix~\ref{app:ReNr}.  We investigate how the convergence of the BET simulation relies on the ensemble size and the time step in Appendix~\ref{app:Convergence}. Finally, we discuss the possibility of apply the BET in different platforms Appendix~\ref{app:Application}.

}

\section{Method}
\label{sec:Method}

{In the Brownian oscillator model, the interaction between the system and the bath can be described by the total Hamiltonian $H=H_{\rm S}+H_{\rm B}+H_{\rm SB}$ \cite{Breuer2002} with}
\begin{eqnarray}
H_{\rm S}&=&\sum_{j} \varepsilon_{j} |j\rangle \langle j|+\sum_{j\neq i} J_{ji} |j\rangle \langle i|,\\
H_{\rm B}&=&\sum_{j,k} \omega_{jk} a_{jk}^\dagger a_{jk},\\
H_{\rm SB}&=&\sum_{j,k} g_{jk} |j\rangle \langle j| (a_{jk}^\dagger+a_{jk} ),
\end{eqnarray}
where $\varepsilon_{j}$ is the site energy of $|j\rangle$, $J_{ji}$ is the coupling constant between the state $|j\rangle$ and $|i\rangle$, $a_{jk}^\dagger$ ($a_{jk}$) is the creation (annihilation) operator for the $k$th mode in the bath of the state $|j\rangle$ with frequency $\omega_{jk}$. The system-bath interaction is given by the spectral density {$\mathcal{J}_j(\omega)=\sum_k g_{jk}^2 \delta(\omega-\omega_k)$} \cite{Breuer2002}. Although a dephasing-form system-bath interaction is adopted here for illustration, the BET framework is also capable of effectively reproducing relaxation dynamics \cite{Soare2014NP,Soare2014PRA,Chen2025JCP}.

Based on the BET \cite{Soare2014NP,Soare2014PRA,Wang2018NPJQI,zhang2021FOP}, we can simulate the dynamics of the open quantum system by introducing a {noise} Hamiltonian of the form
\begin{eqnarray}
H_{\rm PDN}(t)&=& \sum_j B_j(t)|j\rangle \langle j|,\\
B_j(t)&=&A_{j} \sum_{n=1}^{n_c}\omega_n F_j(\omega_n)\cos(\omega_n t + \phi_n^{(j)}),
\end{eqnarray}
to the system Hamiltonian $H_{\rm S}$, such that the quantum-simulation Hamiltonian reads
\begin{equation}
H_{\rm QS}(t)=H_{\rm S}+H_{\rm PDN}(t).
\end{equation} 
Here $B_j(t)$ is a time-dependent stochastic variable mimicking the effect of the environment of the $j$th state. $A_{j}$ is a global scaling factor which characterizes the strength of the noise. The random phases are given by $\{\phi_n^{(j)}|n=1,2\cdots n_c\}$ with each $\phi_n^{(j)}$ uniformly distributed in $[0,2\pi)$. The corresponding mode frequencies are $\omega_n=n\omega_0$, where $\omega_0$ is the base frequency and $n_c \omega_0$ is the cutoff frequency. $F_j(\omega)$ characterizes the shape of the noise's correlation function in the frequency domain.

The two-time correlation function of $B_j(t)$ is
\begin{equation}\label{eq:BB}
\langle B_j(t+\tau)B_j(\tau) \rangle=\frac{A_{j}^2}{2}\sum_{n=1}^{n_c}[\omega_n F_j(\omega_n)]^2 \cos(\omega_n t),
\end{equation}
where $\langle \cdots \rangle$ is the average over the ensemble.
%{The power-spectral density $S_j(\omega)=\int {\rm d}\tau \langle B_j(t+\tau)B_j(t) \rangle_{\rm ensemble} \exp(-i\omega\tau)$ is the Fourier transform of the correlation function, i.e.,
%\textcolor{red}{\begin{equation}
%S_j(\omega)=\frac{\pi}{2}A_j^2 \sum_{n=1}^{n_c} \omega_n^2 F_j^2(\omega_n)[\delta(\omega-\omega_n)+\delta(\omega+\omega_n)].
%\end{equation}}}
%In order to exactly simulate the open quantum dynamics, we require the decoherence factor and the lineshape function to have the same time dependence, i.e., $\chi(t)={\rm Re}[g(t)]$ \cite{zhang2021FOP}, where
%\textcolor{red}{\begin{eqnarray}
%\chi(t)&=&A_j^2 \sum_{n=1}^{n_c}[F_j(\omega_n)]^2 \sin^2 \frac{\omega_n t}{2},\\
%g(t)&=&\int {\rm d}\omega \frac{\mathcal{J}(\omega)}{\omega^2}[(1-\cos \omega t)\coth \frac{\beta \omega}{2}+i(\sin \omega t -\omega t)],\nonumber
%\end{eqnarray}}
As shown in Appendix~\ref{app:TCF}, the TCF of the system can be expressed in terms of the spectral density $\mathcal{J}_j(\omega)$ as
\begin{equation}\label{eq:TCF}
C_j(t)=\int_0^{\infty} {\rm d}\omega \mathcal{J}_j(\omega)\left[\coth \left(\frac{\beta \omega}{2}\right)\cos (\omega t)-i\sin (\omega t)\right],
\end{equation}
with $\beta = 1/(k_B T_{\mathrm{env}})$, where $k_B$ is the Boltzmann constant and $T_{\mathrm{env}}$ denotes the temperature of the environment. 
In the high-temperature regime, the TCF becomes real and the simulation of open quantum dynamics requires $\langle B_j(t+\tau)B_j(\tau) \rangle={\rm Re}[C_j(t)]$ \cite{zhang2021FOP}. Equivalently, the decoherence factor $\chi_j(t)$ equals the real part of the lineshape function $g_j(t)$, i.e., $\chi_j(t)={\rm Re}[g_j(t)]$ \cite{zhang2021FOP}, where
\begin{align}
\chi_j(t)=&A_j^2 \sum_{n=1}^{n_c}[F_j(\omega_n)]^2 \sin^2 \frac{\omega_n t}{2},\\
g_j(t)=&\int {\rm d}\omega \frac{\mathcal{J}_j(\omega)}{\omega^2}[(1-\cos \omega t)\coth \frac{\beta \omega}{2}\!+\!i(\sin \omega t \!-\!\omega t)].
\end{align}
With $F_j(\omega)$ determined from the Fourier transform of ${\rm Re}[C_j(t)]$ and a global scaling factor $A_j$, the target TCF can be reconstructed. By further selecting appropriate the base frequency $\omega_0$ and the cutoff $n_c$, the open quantum dynamics governed by $H$ can be faithfully reproduced by an ensemble average over $H_{\rm QS}(t)$. 
Further discussions on the types and controllability of environmental noise realizable on various experimental
platforms are provided in Appendix~\ref{app:Application}.

\section{BET-based simulation of 2DS: Four-level chiral system}
\label{sec:Chiral}

\begin{figure}[htp]
\includegraphics[width=8.5cm]{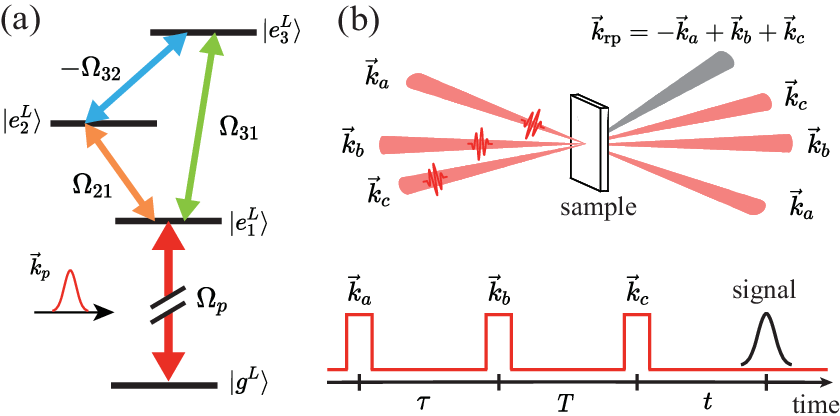}\caption{Schematic of 2DS and its application for enantiodetection of chiral molecules. (a) The four-level structure in a left-handed chiral molecule with three control fields applied for distinguishing the two chiral forms. The difference between the two chiral forms lies in the fact that the transition dipole moments between $|e_2^{L} \rangle\leftrightarrow|e_3^{L} \rangle$ and $|e_2^{R} \rangle\leftrightarrow|e_3^{R} \rangle$ are opposite in sign, resulting in $\Omega_{32}^{L}=-\Omega_{32}^{R}=-\Omega_{32}$. The incident probe pulses with wave vectors $\vec{k}_p~(p=a,b,c)$ are applied to induce the transition between $|g^{\alpha}\rangle$ and $|e_1^{\alpha}\rangle$ with Rabi frequency $\Omega_p$. (b) Schematic illustration of 2DS. The detection pulses employ the square-wave approximation, and the curve represents the signal.
% Double-side Feynman diagrams for (c) stimulated emission process and (d) ground-state bleaching process.
\label{fig:system}}
\end{figure}

{As a demonstration, we simulate the 2DS of a four-level system consisting of a cyclic three-level subsystem and a ground state for enantiodetection of chiral molecules \cite{cai2022PRL}, where three electric-dipole transitions are resonantly driven by three electromagnetic driving fields, as depicted in Fig.~\ref{fig:system}(a).} In the interaction picture with respect to {$H_0^\alpha=\sum_{j=1}^3 e_j| e_j^\alpha \rangle \langle e_j^\alpha |$}, the effective Hamiltonian reads
\begin{align}
	H_{\rm eff}^{\alpha} \!=&\!  H_{\rm I}^{\alpha}
	 \!+\! \sum_{j,k} [\omega_{ik} a_{jk}^\dagger a_{jk}  \! + \!g_{jk} |e_j^\alpha\rangle \langle e_j^\alpha|(a_{jk}^\dagger \!+ \!a_{jk})] ,\\
H_{\rm I}^{\alpha} \!=& \!\Omega_{21}^{\alpha}| e_2^\alpha \rangle \langle e_1^\alpha |\!+\!\Omega_{31}^{\alpha}| e_3^\alpha \rangle \langle e_1^\alpha |\!+\!\Omega_{32}^{\alpha}| e_3^\alpha \rangle \langle e_2^\alpha |\!+\!{\rm h.c.},\!
\end{align}
where $\alpha=L,R$ represents left- and right-handed chiral molecules, respectively, with $\Omega_{21}^{L}=\Omega_{21}^{R}=\Omega_{21}$, $\Omega_{31}^{L}=\Omega_{31}^{R}=\Omega_{31}$, $\Omega_{32}^{L}=-\Omega_{32}^{R}=-\Omega_{32}$.
{Time-dependent stochastic variables $B_j(t)$'s are used to simulate the impact of the environment and thus the simulation Hamiltonian reads
{\begin{equation}
		H_{\rm QS}^{\alpha}(t) = H_{\rm I}^{\alpha}+\sum_j B_j(t)|e_j^{\alpha}\rangle \langle e_j^{\alpha}| .
\end{equation}}
%In the BET, the reduced density matrix of the system evolves according to $\rho^{\alpha}_{\rm sys}(t) = \langle U_{\rm QS}^{\alpha}(t,t_0)\rho^{\alpha}_{\rm sys}(t_0)U_{\rm QS}^{\alpha \dagger}(t,t_0)\rangle_{\rm ens}$ with the evolution operator $U_{\rm QS}^{\alpha}(t,t_0)=\mathcal{T} \exp[-i\int_{t_0}^t {\rm d} \tau H_{\rm QS}^{\alpha}(\tau)]$. The average is taken over an ensemble of $N$ Hamiltonians $H_{\rm QS}^{\alpha}(t)$'s each characterised by a distinct set of phases $\phi_n^{(j)}$'s.

{In 2DS, three pulses $\vec{E}_a$, $\vec{E}_b$, and $\vec{E}_c$ with incident wave vectors $\vec{k}_p~(p=a,b,c)$, sequentially interact with the sample to generate a nonlinear signal field $\vec{E}_s$, as illustrated in Fig.~\ref{fig:system}(b).}
In the interaction picture, {the Hamiltonian during a pulse is written as}
\begin{equation}
V_p^\alpha(t) = \Omega_p(t) e^{i\vec{k}_p \cdot \vec{r}}\mu_+^{\alpha}  + {\rm h.c.},
\end{equation}
where $\mu_+^{\alpha}=|e_1^\alpha \rangle \langle g^\alpha |=(\mu_-^{\alpha})^\dagger$ is the dipole operator.
The sample interacts initially with the first pulse along $\vec{k}_a$, followed by an interaction with the second pulse along $\vec{k}_b$ after the coherence time $\tau$. Subsequently, after the population time $T$, the sample interacts with the third pulse along $\vec{k}_c$. The signals emitted {from} the sample {are} measured after the signal time $t$. In typical experiments, the rephasing and non-rephasing signals are measured along the directions $\vec{k}_{\rm rp}=-\vec{k}_a+\vec{k}_b+\vec{k}_c$ and $\vec{k}_{\rm nr}=\vec{k}_a-\vec{k}_b+\vec{k}_c$, respectively. 
 %{From these emitted signals after all the evolutions, the 2DS is obtained.} 

%Finally, in the interaction picture, the wave function is given as
%\begin{eqnarray}
%|\psi_I^\alpha(\tau,T,t)\rangle &=& U_{\rm QS}^\alpha(t+T+\tau,T+\tau)U_{cI}^\alpha
% U_{\rm QS}^\alpha(T+\tau,\tau)\nonumber\\& &U_{bI}^\alpha U_{\rm QS}^\alpha(\tau,0)U_{aI}^\alpha |\psi_0^\alpha \rangle ,
%\end{eqnarray}
%where the initial state $|\psi_0^\alpha \rangle$ is taken as the ground state $|g^\alpha \rangle$. 

{Given that the initial state of the system is the thermal-equilibrium state and that the energy gap between the ground state and the excited states is large, the system can be regarded as being in the ground state. Therefore, the initial density matrix $\rho_0^{\alpha}$ can simply be taken as the pure state $|g^\alpha \rangle$. In the BET, there are multiple pathways governed by the ensemble of Hamiltonians, and each pathway will process the interactions with the pulses and the free evolutions, resulting in a set of final states. In the interaction picture, the final state of each pathway is given as
\begin{align}
|\psi_I^\alpha(\tau,T,t)\rangle=& U_{\rm QS}^\alpha(t+T+\tau,T+\tau)U_{cI}^\alpha
 U_{\rm QS}^\alpha(T+\tau,\tau)\nonumber\\
& \times U_{bI}^\alpha U_{\rm QS}^\alpha(\tau,0)U_{aI}^\alpha |g_0^\alpha \rangle .
\end{align}}
{The evolution operator in the presence of the pulse is $U_{pI}^\alpha=\mathcal{T}\exp[-i \int_{0}^{\delta t_p} {\rm d} s V_p^\alpha(s) ]$, which {is} approximated to {the} first order as $U_{pI}^\alpha \approx I-iV_p^\alpha\delta t_p$, where $V_p^\alpha = \Omega_p \exp (i\vec{k}_p \cdot \vec{r}) \mu_+^{\alpha} + {\rm h.c.}$ Here, we use the square-pulse approximation and assume that $\Omega_p\delta t\ll1$. On this basis, the third-order polarization signals in the rephasing and the non-rephasing directions can be conveniently calculated by the following equations
\begin{align} 
P_{\rm rp}^\alpha(t,T,\tau)=&\langle\textrm{Tr} \left(\mu_-^{\alpha}\mathcal{G}^{\alpha}_{t}\{\mu_+^{\alpha}\mathcal{G}^{\alpha}_{T}[\mathcal{G}^{\alpha}_{\tau}(\rho_0^{\alpha}\mu_-^{\alpha})\mu_+^{\alpha}]\}\right)\rangle+\nonumber\\ %\Rightarrow {\rm GSB}
 & \langle\textrm{Tr} \left( \mu_-^{\alpha}\mathcal{G}^{\alpha}_{t}\{\mathcal{G}^{\alpha}_{T}[\mu_+^{\alpha}\mathcal{G}^{\alpha}_{\tau}(\rho_0^{\alpha}\mu_-^{\alpha})]\mu_+^{\alpha}\} \right)\rangle, \label{eq:Prpt}\\%\Rightarrow {\rm SE}
P_{\rm nr}^\alpha(t,T,\tau) =&\langle\textrm{Tr} \left( \mu_-^{\alpha}\mathcal{G}^{\alpha}_{t}\{\mu_+^{\alpha}\mathcal{G}^{\alpha}_{T}[\mu_-^{\alpha}\mathcal{G}^{\alpha}_{\tau}(\mu_+^{\alpha}\rho_0^{\alpha})]\}\right)\rangle +\nonumber\\ %\Rightarrow {\rm GSB}
 & \langle\textrm{Tr} \left( \mu_-^{\alpha}\mathcal{G}^{\alpha}_{t}\{\mathcal{G}^{\alpha}_{T}[\mathcal{G}^{\alpha}_{\tau}(\mu_+^{\alpha}\rho_0^{\alpha})\mu_-^{\alpha}]\mu_+^{\alpha}\}\right)\rangle.\label{eq:Pnrt} %\Rightarrow {\rm SE}
\end{align}
Note that the leftmost $\mu_-^{\alpha}$ is used to compute the polarization signal while the other three describe the interaction with a pulse. The superoperators $\mathcal{G}^{\alpha}_{\tau}$, $\mathcal{G}^{\alpha}_{T}$ and $\mathcal{G}^{\alpha}_{t}$ denote the free evolution of the system's reduced density matrix governed by the time evolution operator $U_{\rm QS}^\alpha(\tau,0)$, $U_{\rm QS}^\alpha(T+\tau,\tau)$ and $U_{\rm QS}^\alpha(t+T+\tau,T+\tau)$, respectively, in the absence of external pulses. 
The final signal in the time domain is obtained by the ensemble average of the time-domain signals of all pathways. {Note that the first terms in Eq.~(\ref{eq:Prpt}) and (\ref{eq:Pnrt}) correspond to the ground-state bleaching, while the second terms correspond to the stimulated emission, as illustrated by the double-side Feynman diagrams in Fig.~\ref{fig:Feynman_2level}.}
 The 2DS is obtained as the real part of the double Fourier transform of the time-domain signal with respect to the coherence and detection times, $\tau$ and $t$, i.e.,
$\tilde{P}_{x}^\alpha(\omega_t,T,\omega_\tau) \equiv {\rm Re}\left(\mathcal{F}_{\tau}\left\{\mathcal{F}_{t}[P_{ x}^\alpha(t,T,\tau)]\right\}\right)$, $x\in \{{\rm rp, nr}\}$. A detailed derivation of the rephasing and non-rephasing signal is presented in Appendix~\ref{app:ReNr}.
\begin{figure}
\includegraphics[width=8.5cm]{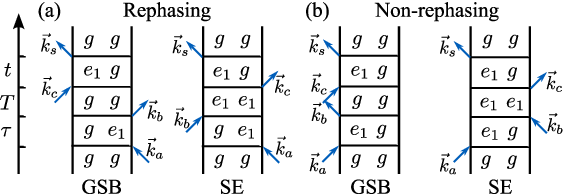}
\caption{Double-sided Feynman diagrams for the (a) rephasing and (b) non-rephasing signals. The arrows represent the incident probe pulses and the signals.} \label{fig:Feynman_2level}
\end{figure}

\begin{figure}
\includegraphics[width=8.5cm]{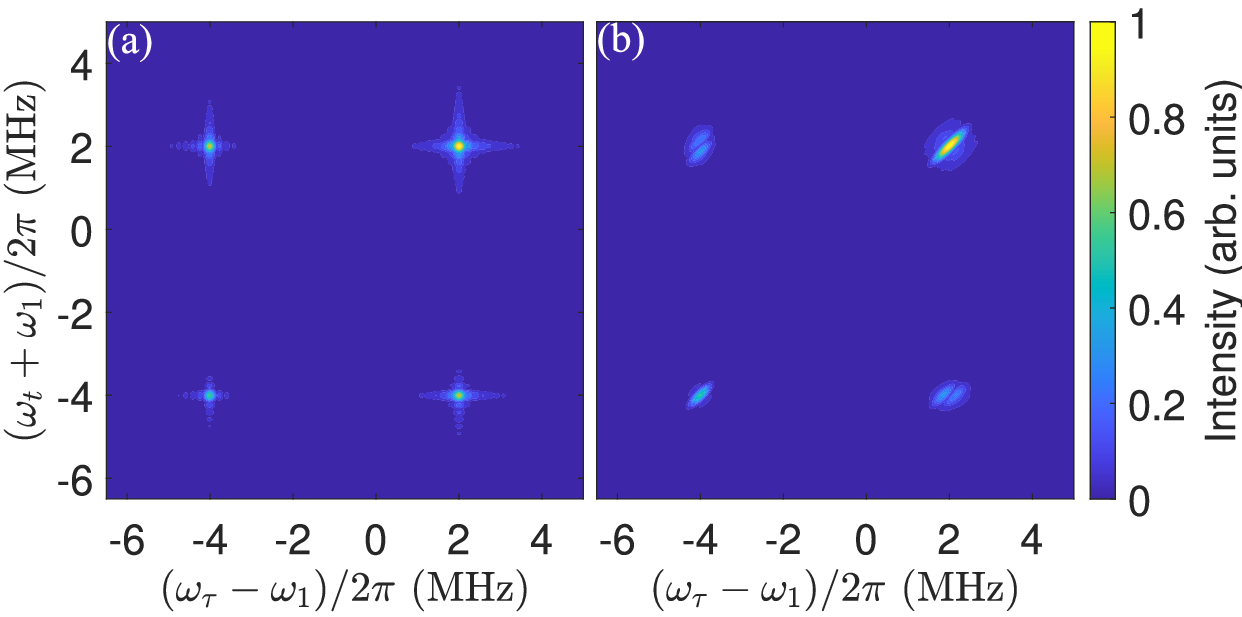}
\caption{The rephasing signals of the 2DS simulated by the BET with (left) linear $\chi(t)$, (right) nonlinear $\chi(t)$ at population time $T=0$. In the linear case, the parameters are $\gamma/2\pi=0.1$~MHz, $A_j=8$~MHz, $n_c=750$, $\omega_0/2\pi=1$~kHz, while in the nonlinear case, the parameters are $A_j=350$~MHz, $n_c=750$, $\omega_0/2\pi=25$~Hz. For both cases, the ensemble size is $N=6000$. }
\label{fig:ComparisonT0}
\end{figure}

 In the demonstration on enantiodetection of chiral molecules using 2DS, 1,2-propanediol with equal Rabi frequencies $\Omega_{21}=\Omega_{31}=\Omega_{32}=\Omega$ is considered \cite{cai2022PRL}. In this case, two of the three eigenstates of $H_I^\alpha$ are degenerate, and the eigenvalues of the non-degenerate state is twice that of the degenerate states with a negative sign, i.e., $E_1^\alpha=E_2^\alpha=-E_3^\alpha/2$. Therefore, there are four peaks in the obtained 2DS, located at $(-2\Omega,\Omega)$, $(\Omega,\Omega)$, $(-2\Omega,-2\Omega)$, $(\Omega,-2\Omega)$ in the case of left-handed chiral molecules, as can be observed in Fig.~\ref{fig:ComparisonT0}. We present the result by the quantum simulation with $\chi(t)=\gamma t$ in Fig.~\ref{fig:ComparisonT0}(a). The TCF takes the form $\delta(t)$ in this condition, indicating the system is Markovian. In the calculation, we take the Rabi frequency $\Omega/2\pi=2$~MHz and the pure-dephasing rate $\gamma/2\pi=0.1$~MHz. Given that the decoherence factor $\chi(t)$ can only be considered linear if the decoherence is Markovian \cite{Breuer2002}, we consider a more-general situation when $\chi(t)$ possesses a quadratic form in short times, while it scales linear with $t$ in the long run. The result is quite different from the linear situation, as shown in Fig.~\ref{fig:ComparisonT0}(b). Both the diagonal and the cross peaks are elongated diagonally, implying a correlation between the two frequency axes.
}

\section{Center-line Slope Theory}
\label{sec:CLS}

2DS can be used to detect the information of both the energy structure and the interaction between the environment and the system. One of the methods for obtaining the system-bath coupling is the CLS theory, which {indicates} that the TCF can be obtained from the slope of the center line of the 2DS \cite{Kwak2007JCP,Rosenfeld2011S}. Due to assumptions made in the derivation, such as the short-time approximation, the CLS theory results in equal CLSs for both the rephasing and non-rephasing signals when the TCF is real, cf. Appendix~\ref{app:cls}, which is not practical in the visible regime \cite{Sun2023AQT}. Since the BET allows for constructing an environment with a specific TCF, we assess the CLS method by comparing the CLS from the 2DS with the preset TCF.

 We adopt the TCF of exponential form as \cite{Kwak2007JCP}
 \begin{equation}\label{eq:Ct}
 C(t)=\Delta^2 e^{-|t|/\tau_c},
 \end{equation}
 where $\Delta$ is the fluctuation amplitude and $\tau_c$ is the correlation time. {Performing a double integration of TCF yields the lineshape function, i.e., $g(t)=\int_0^t dt_1 \int_0^{t_1} dt_2 C(t_2)$, which evaluates to $g(t)=\Delta^2 \tau_c^2(e^{-t/\tau_c}+t/\tau_c-1)$.} Given that $C(t)$ is an even function, it {is} expanded into Fourier series as $\sum_{n=0}d_n\cos(\omega_nt)$. By appropriately selecting the cutoff frequency $n_c\omega_0$ and the base frequency $\omega_0$, Eq.~(\ref{eq:BB}) can {accurately} reproduce the TCF in Eq.~(\ref{eq:Ct}), cf. Appendix~\ref{app:Convergence}.
The TCF is extracted from { the CLS of} the absorptive 2DS, which is obtained by the sum of the real parts of the signals in rephasing and non-rephasing directions, i.e., {$\tilde{P}_{\rm abs}^\alpha(\omega_\tau,T,\omega_t)={\rm Re}[\tilde{P}_{\rm rp }^\alpha(\omega_\tau,T,\omega_t)+\tilde{P}_{\rm nr}^\alpha(\omega_\tau,T,\omega_t)]$.} The details for the derivation are provided in Appendix~\ref{app:ReNr}.
%The TCF is extracted from { the CLS of} the absorptive 2DS, which is obtained by the combination of the real parts of the {signals in rephasing and non-rephasing directions,} i.e., {$\tilde{P}_{\rm abs}^\alpha(\omega_\tau,T,\omega_t)={\rm Re}[\tilde{P}_{\rm rp }^\alpha(\omega_\tau,T,\omega_t)+\tilde{P}_{\rm nr}^\alpha(\omega_\tau,T,\omega_t)]$.} The details can be referred to in the Supporting Information.
%{\color{cyan}??In the interaction picture, the wave function is given as
%\begin{eqnarray}
%|\psi_I^\alpha(\tau,T,t)\rangle &=& U_{\rm QS}^\alpha(t+T+\tau,T+\tau)U_{cI}^\alpha
% U_{\rm QS}^\alpha(T+\tau,\tau)\nonumber\\& &U_{bI}^\alpha U_{\rm QS}^\alpha(\tau,0)U_{aI}^\alpha |\psi_0^\alpha \rangle ,
%\end{eqnarray}
%where the initial state $|\psi_0^\alpha \rangle$ is taken as the ground state $|g^\alpha \rangle$. }
 \begin{figure*}[htp]
\includegraphics[bb=50 0 800 278, width=16cm]{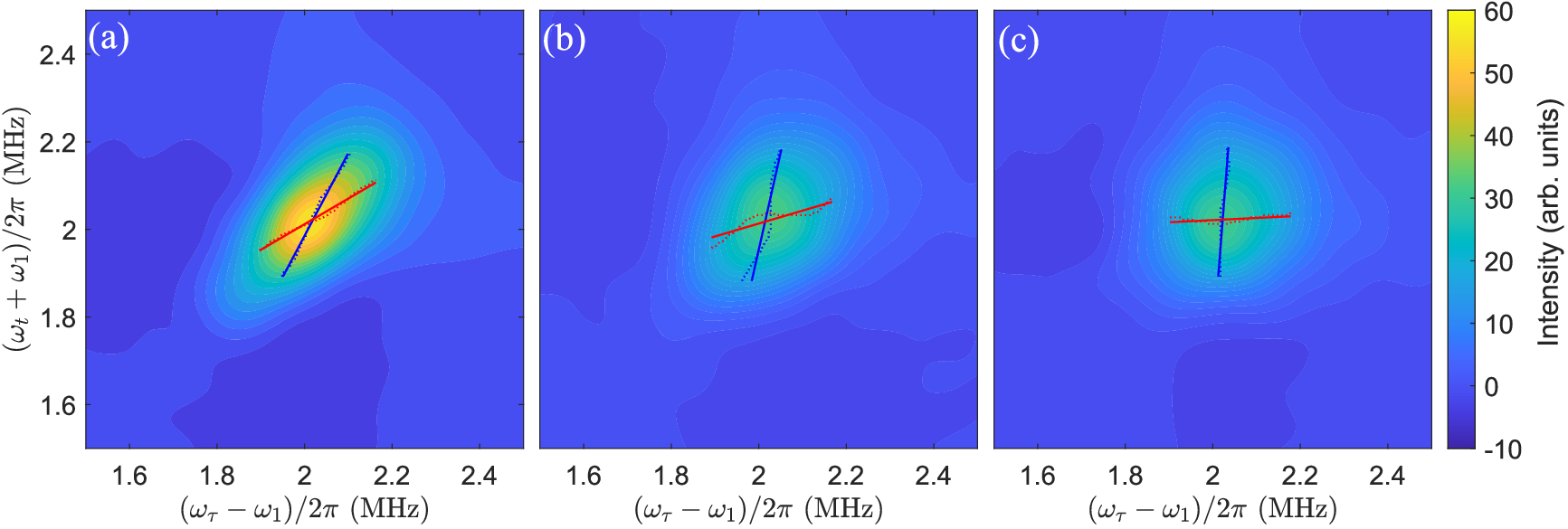} %
\caption{The top-right diagonal peak in the absorptive 2DS of pure left-handed chiral molecules simulated by the BET at population time $T=0, 4, 8~{\rm \mu s}$. The red (blue) dotted lines are the center lines for $\omega_\tau$ ($\omega_t$). The red (blue) solid lines are the least-squares fitting of the center lines for $\omega_\tau$ ($\omega_t$). The parameters of the TCF are $\Delta=1$~MHz and $\tau_c=4~\mu$s, while the other parameters are $A_j=1$~MHz, $n_c=3.6\times10^4$, $\omega_0/2\pi=125$~Hz and the ensemble size is $N=6000$.
 \label{fig:rephasing2d}}
\end{figure*}

\begin{figure}[htp]
\includegraphics[bb=0 0 395 278, width=8.5cm]{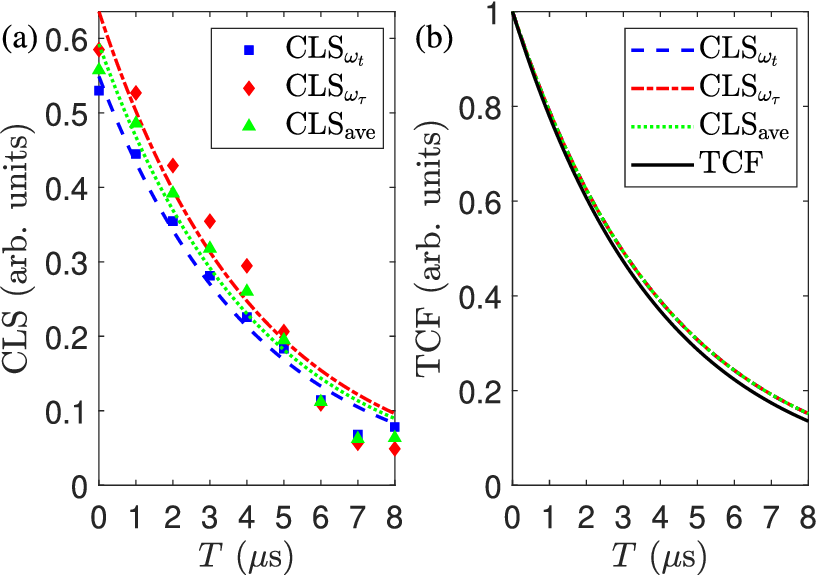}\caption{Comparison of the CLS with TCF. (a) The dependence of the CLS on the population time $T$. The red diamonds, blue squares and green triangle are the CLS with respect to $\omega_\tau$ and $\omega_{t}$, and their average. The red dashed-dotted line, blue dashed line, and the green dotted line are the corresponding fitted results. (b) Comparison of the normalized fitted results with the TCF plotted as the black solid line. \label{fig:rephasingcls}}
\end{figure}

 {Figure~\ref{fig:rephasing2d} demonstrates} the time evolution of the top-right diagonal peak in the absorptive 2DS. The center lines are depicted by red (blue) lines, whose slopes are calculated with respect to the $\omega_\tau$ ($\omega_t$) axis. In order to obtain the CLS effectively, the center lines are restricted to the {full width at the half maximum} of the peaks and are fitted by a straight line using the least squares method. As the population time $T$ increases, the height of the peak decreases, and the CLSs with respect to $\omega_t$ and $\omega_\tau$ decrease and tend to zero.
 As depicted in Fig.~\ref{fig:rephasingcls}(a) and (b), {we fit each of the two CLS types} and their average by an exponential function of $a \exp(-t/\tau_c)$, and compare the normalized results with the preset TCF. The evolution of CLSs over time can be well described by an exponential decay, and the normalized results are consistent with the preset TCF. However, for the rephasing and non-rephasing 2DS, the CLSs deviate significantly from the preset TCF, as shown in Figs.~\ref{fig:rephasingfig048}--\ref{fig:nonrephasingfig4} of Appendix~\ref{app:ReNr}. The CLS obtained from the rephasing 2DS generally presents a monotonic behavior, but with a decay rate much lower than the preset one, while the CLS obtained from the non-rephasing 2DS exhibits a non-monotonic dependence on $T$.
  %Nevertheless, the relative amplitudes of the rephasing and non-rephasing 2DS vary with $T$, which leads to an overall exponential trend consistent with the preset TCF in the absorptive 2DS.}
 %If the TCF is more complex, the result of CLS method is worse.
 %the CLS decays faster than the TCF. The decay time of TCF is set as $4~{\rm \mu s}$, whereas the decay time of the ${\rm CLS_{ave}}$ in the 2DS is $1.612~{\rm \mu s}$. Additionally, as time progresses, the CLS will exhibit negative values ???.

\section{Simulation for RDC Dissolved in Chloroform}
\label{sec:RDC}

To further validate our simulation approach, we perform the simulation of the 2DS for RDC dissolved in chloroform. This system consists of six energy levels and has been previously characterized by both experimental and theoretical results \cite{Demirdoven2002PRL,Khalil2003JPCA}. The simulation parameters are taken therein, and the result is shown in Fig.~\ref{fig:RDC}.

\begin{figure}
\centering
\includegraphics[width=8.5cm]{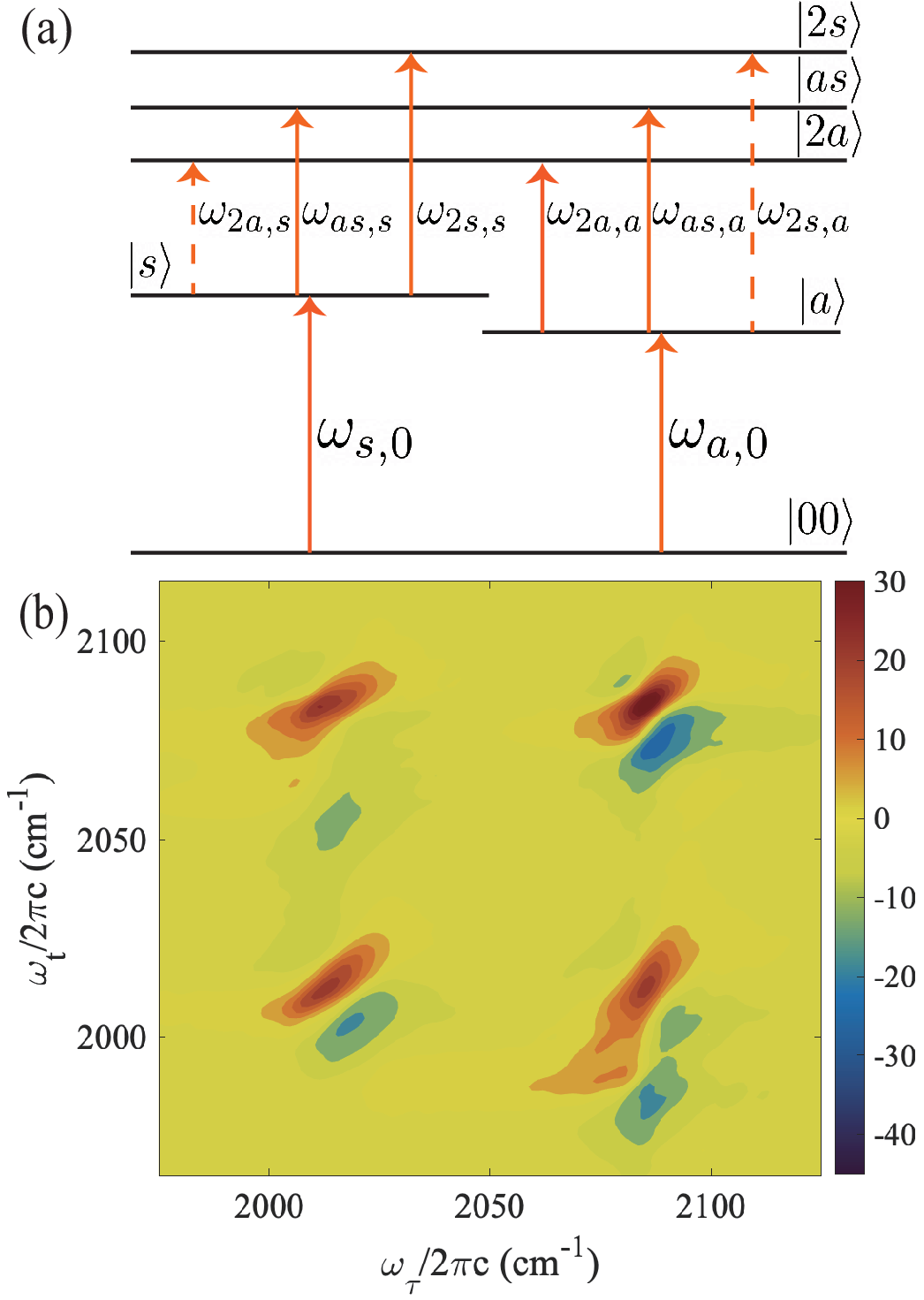}
\caption{(a) The energy-level structure of the RDC. The solid (dashed) arrows represent the allowed (forbidden) transitions. (b) The absorptive 2DS at $T=0$ obtained by the BET-based quantum-simulation approach.}
\label{fig:RDC}
\end{figure}

Figure~\ref{fig:RDC}(a) demonstrates the energy-level structure of the RDC, where the solid arrows denote the allowed transitions and the dashed arrows denote the forbidden ones. The system includes a ground vibrational state $\vert 00\rangle$, two one-quantum states $\vert a\rangle$ and $\vert s\rangle$, and three two-quantum states $\vert 2a \rangle$, $\vert 2s\rangle$, and $\vert as\rangle$. The transition frequencies between these states are respectively $\omega_{a,0}=2014~\textrm{cm}^{-1}$, $\omega_{s,0}=2085~\textrm{cm}^{-1}$, $\omega_{2a,a}=2001~\textrm{cm}^{-1}$, $\omega_{2s,s}=2073~\textrm{cm}^{-1}$, $\omega_{as,a}=2058~\textrm{cm}^{-1}$, $\omega_{as,s}=1989~\textrm{cm}^{-1}$,
$\omega_{2a,s}=1932~\textrm{cm}^{-1}$, and $\omega_{2s,a}=2142~\textrm{cm}^{-1}$ \cite{Golonzka2001}. 
The non-vanishing elements in electric-dipole operator are $\mu_{a,0}=1.05\mu_{s,0}, \mu_{as,s}=\mu_{a,0}, \mu_{as,a}=\mu_{s,0}, \mu_{aa,a}=1.48\mu_{s,0}, \mu_{ss,s}=1.41\mu_{s,0}, \mu_{ss,a}=\mu_{aa,s}=0.13\mu_{s,0}$ and $\mu_{s,0}=1$ \cite{Golonzka2001}. In addition, the evolution times $\tau$ and $t$ vary from 0 to 5~ps, with a time step of $5~\rm{fs}$. The probe pulses have the same spectral bandwidth, which are wide enough to excite all the vibrational transitions. 

%We consider the correlation effects by adding the fluctuations to the original transition frequencies. 
The TCF for one-quantum states is given by
\begin{align}
C_{p,q}(t)&=\langle \delta \omega_{p,0}(t)\delta \omega_{q,0}(0) \rangle \nonumber \\ &=\rho_{qp}\sigma_{pp}\sigma_{qq} e^{-|t|/\tau_{pq}},~(p,q=a,s),
\end{align}
where $ \delta \omega_{p,0}$ and $ \delta \omega_{q,0}$  are the fluctuations of transition frequencies for one-quantum states. $\sigma_{pp}$ and $\sigma_{qq}$ are the amplitudes for the fluctuations, while $\tau_{pq}$ is the correlation time. $\rho_{qp}$ characterizes the statistical interdependence between $ \delta \omega_{p,0}$ and $ \delta \omega_{q,0}$, which can take values between $-1$ and $+1$. For auto-correlation, $p=q$ and $\rho_{pp}=1$, while for cross-correlation, $p \neq q$. Parameters of the TCF for the one-quantum states are $\sigma_{aa}=7.4~\rm{cm^{-1}}$, $\sigma_{ss}=5.2~\rm{cm^{-1}}$, $\tau_{aa}=2.0~\rm{ps}$, $\tau_{ss}=2.0~\rm{ps}$, $\tau_{as}=1.2~\rm{ps}$, and $\rho_{as}=0.9$ \cite{Khalil2003JPCA,Demirdoven2002PRL}. 

In addition to the auto- and cross-correlations of the one-quantum states, the auto-correlations of the two-quantum states are also considered. Their TCFs are constructed from the one-quantum correlations and are given as
\begin{align}
C_{pq,pq}(t)&=\langle \delta \omega_{pq,0}(t)\delta \omega_{pq,0}(0) \rangle \nonumber\\ &=\eta^2(\sigma_{pp}+\sigma_{qq})^2 e^{-|t|/\sqrt{\tau_{pp}\tau_{qq}}},\nonumber\\
&\quad(p,q=a,s; pq=2a,2s,as).
\end{align}
Here, $\delta\omega_{pq,0}(t)$  denotes the fluctuation of two-quantum states. Since these fluctuations are not fully correlated with those of the one-quantum states, a scaling factor 
$\eta=0.1$ is introduced \cite{Tominaga1997}. The cross-correlations among the two-quantum states, and between the one- and two-quantum states are neglected.

The elements in the noise Hamiltonian of the BET are obtained from the TCF. When taking the cross-correlation into account, $B_j(t)$ would become more complex, and can be divided into two parts: one associated with the cross-correlation, and the other corresponding to the difference between the auto-correlation and the cross-correlation. Accordingly, $B_j(t)=\beta_j^{(\rm{cr})}(t)+\beta_j^{(\rm{ac})}(t),~(j=a,s)$, which satisfies TCFs for both the auto- and cross-correlations. Each part of $B_j(t)$ is written as
\begin{align}
\beta_j^{(\rm{cr})}(t) =\sum_{n=1}^{n_c}& \omega_n \cdot
\frac{1}{n}\sqrt{ \frac{4}{\pi\omega_0} \left(
    \frac{\rho_{as}\sigma_{aa}\sigma_{ss} \tau_{as}}{1 + \omega_n^2 \tau_{as}^2}
\right)}\nonumber\\
&\times \cos\left(\omega_n t + \phi_n^{(j,{\rm{cr}})}\right),\label{eq:beta_j_cr} \\
\beta_j^{(\rm{ac})}(t) =\sum_{n=1}^{n_c}& \omega_n \cdot
\frac{1}{n}\sqrt{ \frac{4}{\pi\omega_0} \left(
    \frac{\sigma_{jj}^2 \tau_{jj}}{1 +\omega_n^2 \tau_{jj}^2} 
    - 
    \frac{\rho_{as}\sigma_{aa}\sigma_{ss} \tau_{as}}{1 + \omega_n^2 \tau_{as}^2}
\right)}\nonumber\\
&\times \cos\left(\omega_n t + \phi_n^{(j,{\rm{ac}})}\right)\label{eq:beta_j_ac}.
\end{align}

For the two-quantum states, $B_j(t)$ $(j=2a,2s,as)$ with only the auto-correlation taken into account are written as
\begin{align}
B_j(t) =\sum_{n=1}^{n_c}& \omega_n \cdot
\frac{1}{n}\sqrt{ \frac{4}{\pi\omega_0} \left(
    \frac{\eta^2(\sigma_{pp}+\sigma_{qq})^2 \sqrt{\tau_{pp}\tau_{qq}}}{1 +\omega_n^2 \tau_{pp}\tau_{qq}} 
\right)}\nonumber\\
&\times \cos(\omega_n t + \phi_n^{(j)})\label{eq:B_j_2quantum},
\end{align}
where the mapping between $j$ and $(p, q)$ is defined as $j=2a \rightarrow (p,q)=(a,a)$, $j=2s \rightarrow (p,q)=(s,s)$ and $j = as \rightarrow (p,q)=(a,s)$.
The sets of random phases $\{\phi_n^{(a,{\rm{ac}})},(n=1,2,\cdots,n_c)\}$ and $\{\phi_n^{(s,{\rm{ac}})},(n=1,2,\cdots,n_c)\}$ are mutually independent and uniformly distributed in $[0,2\pi)$.
$\{\phi_n^{(2a)},(n=1,2,\cdots,n_c)\}$ and $\{\phi_n^{(2s)},(n=1,2,\cdots,n_c)\}$ for the auto-correlations of $\vert 2a \rangle$ and $\vert 2s \rangle$ are identical to $\{\phi_n^{(a,{\rm{ac}})},(n=1,2,\cdots,n_c)\}$ and $\{\phi_n^{(s,{\rm{ac}})},(n=1,2,\cdots,n_c)\}$, respectively. $\{\phi_n^{(as)},(n=1,2,\cdots,n_c)\}$ here is set to be the average of these two sets. For the cross-correlations of $\vert a \rangle$ and $\vert s \rangle$, the random phases are identical, i.e. $\{\phi_n^{(a,{\rm{cr}})},(n=1,2,\cdots,n_c)\}=\{\phi_n^{(s,{\rm{cr}})},(n=1,2,\cdots,n_c)\}$, and are drawn from another independent set uniformly distributed in $[0,2\pi)$.

Figure~\ref{fig:RDC}(b) is the simulation of absorptive 2DS using the BET-based quantum-simulation approach. To obtain it, the time-domain rephasing and non-rephasing signals are calculated. Then, we apply zero padding to expand the length of the time-domain data to 5000 points per dimension. The absorptive signal is the real part of the summation of the rephasing and non-rephasing signals after the double Fourier transforms.

Our simulation faithfully reproduces the experimental observations in Refs.~\cite{Khalil2003JPCA,Demirdoven2002PRL}. It exhibits the same number, positions, signs, and tilts of all peaks, whose origin can be illustrated as follows. 
After the first pulse, the system is promoted to optical coherences within the one-quantum manifold. Accordingly, along the $\omega_{\tau}$ axis, peaks appear at $\omega_{\tau}=\omega_{a,0},~\omega_{s,0}$. 
After the subsequent interactions, the detected third-order polarization evolves during the detection time $t$ either on the coherences between the ground state and the one-quantum states or on the coherences between the one-quantum states and the two-quantum states. Consequently, the signals at $\omega_{t}=\omega_{a,0},~\omega_{s,0}$ are dominated by the ground-state bleaching, giving rise to the positive peaks, whereas the signals at $\omega_{t}=\omega_{2a,a},~\omega_{2s,s},~\omega_{as,a},~\omega_{as,s}$ are dominated by the excited-state absorption, leading to the negative peaks. 
 Accordingly, the tilt of the positive peaks is primarily governed by the frequency-frequency correlations within the one-quantum manifold. By contrast, the tilt of the negative peaks is dominantly influenced by the correlations associated with the two-quantum states.
 % Accordingly, the correlations between the one-quantum states have the main influence on the tilt of the positive peaks, but not on the negative peaks. Meanwhile, the auto-correlations of the two-quantum states dominantly influence the tilt of the negative peaks.

\section{Conclusion}
\label{sec:Conclusion}

In this work, we develop a BET-based quantum-simulation method to simulate the 2DS of open quantum systems. We simulate the 2DS of a four-level system for the enantiodetection of chiral molecules. Furthermore, we assess the CLS theory using the simulated spectra. The results represent that the preset TCF can be recovered from the absorptive 2DS, whereas the CLS theory fails for the 2DS of the rephasing and non-rephasing signals. In this way, we clarify the application of the CLS theory in practice. To further demonstrate our approach, we apply it to the simulation of the 2DS for the RDC dissolved in chloroform, where the result is consistent with previous experimental observations. Our approach provides an alternative for exploring the dynamics of open quantum systems through the 2DS, and it is particularly suited for structured environments where conventional treatments may become computationally demanding. This approach is expected to enable high-efficiency quantum simulation of the 2DS in experimental platforms, which may explore quantum coherence as a design element for realizing functionalities in chemical and biophysical systems \cite{Scholes2017Nature}.
%Here, we employed a BET to exactly quantum simulate 2DS. Compared with the non-Hermitian Hamiltonian approach, our approach faithfully reproduces the enantiodetection of chiral molecules by the 2DS.We further demonstrate that the TCF can be effectively obtained by applying the CLS theory to the absorptive 2DS, while the CLS theory fails for the rephasing and non-rephasing signals of the 2DS.

\begin{acknowledgments}
We thank stimulating discussions with Y. Song, M.-R. Cai, and X. Leng. This work is supported by the National Natural Science Foundation of China under Grant No.~62461160263, Innovation Program for Quantum Science and Technology under Grant No.~2023ZD0300200,  and
Guangdong Provincial Quantum Science Strategic Initiative under Grant No.~GDZX2505004. F.N. is supported in part by: Nippon Telegraph and Telephone Corporation (NTT) Research,
the Japan Science and Technology Agency (JST) [via
the Quantum Leap Flagship Program (Q-LEAP), and the Moonshot R\&D Grant Number JPMJMS2061], the Asian Office of Aerospace Research and Development
(AOARD) (via Grant No. FA2386-20-1-4069), and
the Office of Naval Research (ONR) (via Grant No.
N62909-23-1-2074).
\end{acknowledgments}

\appendix
\section{TCF of Electronic System Coupling to Harmonic Vibrations}
\label{app:TCF}

In this appendix, we derive the TCF of the bath. In the Brownian-oscillator model, the electronic degrees of freedom are coupled to nuclear motions modeled as independent modes of harmonic ocillators. The Hamiltonian of a two-level system reads \cite{Mukamel1999}
\begin{equation}
H=|g\rangle H_{g}\langle g|+|e\rangle H_{e}\langle e|,
\end{equation}
where
\begin{align}
H_{g} & =\sum_{j}\left(\frac{p_{j}^{2}}{2m_{j}}+\frac{1}{2}m_{j}\omega_{j}^{2}q_{j}^{2}\right),\label{eq:H_g}\\
H_{e} & =\hbar\omega_{{\rm eg}}^{0}+\sum_{j}\left[\frac{p_{j}^{2}}{2m_{j}}+\frac{1}{2}m_{j}\omega_{j}^{2}(q_{j}+d_{j})^{2}\right]\label{eq:H_e}
\end{align}
are the Hamiltonians when the electronic system is in the ground state and the excited state, respectively, the energy gap between the ground state and the excited state is $\hbar\omega_{{\rm eg}}^{0}$,
$p_{j}$, $q_{j}$ and $m_{j}$ represent the momentum, the position,
and the mass of the $j$th nuclear mode, respectively. $d_{j}$ is the displacement which characterizes the system-bath coupling. $q_{j}$
and $p_{j}$ can be rewritten in terms of the creation and annihilation operators, i.e., $a_{j}^{\dagger}$ and $a_{j}$, as
\begin{align}
q_{j} & =\frac{1}{\sqrt{2m_{j}\omega_{j}}}(a_{j}^{\dagger}+a_{j}),\\
p_{j} & =\frac{\sqrt{2m_{j}\omega_{j}}}{2i}(a_{j}-a_{j}^{\dagger}).
\end{align}
Thus, the total Hamiltonian becomes
\begin{align}
H & =\left(\hbar\omega_{eg}^{0}+\sum_{j}\frac{1}{2}m_{j}\omega_{j}^{2}d_{j}^{2}\right)|e\rangle\langle e|+\sum_{j}\omega_{j}a_{j}^{\dagger}a_{j}\nonumber\\
&+|e\rangle\langle e|\sum_{j}g_{j}(a_{j}^{\dagger}+a_{j})+\frac{1}{2}\sum_{j}\omega_{j},
\end{align}
with the coupling strength $g_{j} =d_{j}\omega_{j}\sqrt{m_{j}\omega_{j}/2}$.

Initially, the bath modes are assumed to be at the thermal equilibrium while the electronic system is in the ground state, i.e.,
\begin{equation}
\rho_{0}=|g\rangle\langle g|\otimes\frac{e^{-\beta H_{g}}}{{\rm Tr}(e^{-\beta H_{g}})},\label{eq:rhoini}
\end{equation}
with $\beta\equiv1/k_{B}T$. Hereafter, we redefine
\begin{equation}
H_g=\sum_l\hbar\omega_la_l^\dagger a_l
\end{equation}
for simplicity.

According to Eqs.~ (\ref{eq:H_g}) and (\ref{eq:H_e}), the electronic energy gap is 
\begin{equation}
M(t)=\sum_{j}m_j\omega_j^2 d_j q_j.
\end{equation}
The TCF of the electronic energy gap is
\begin{equation}\label{eq:Ctxi}
C(t)=\langle M(t)M(0)\rho_{0}\rangle=\sum_{j}\xi_{j}^{2}c_{j}(t),
\end{equation}
with
\begin{align}
\xi & \equiv m_{j}\omega_{j}^{2}d_{j},\\
c_{j}(t) & \equiv\langle q_{j}(t)q_{j}(0)\rho_{0}\rangle\label{eq:Cjt},
\end{align}
which has a symmetry
\begin{equation}
c_{j}(-t)\equiv\langle q_{j}(-t)q_{j}(0)\rho_{0}\rangle=\langle q_{j}(0)q_{j}(t)\rho_{0}\rangle=c_{j}^{*}(t). \label{eq:symmetry}
\end{equation}
We can separate $c_{j}(t)$ into its real and imaginary parts as
\begin{align}
c_{j}(t) & =c_{j}'(t)+ic_{j}''(t),
\end{align}
where
\begin{align}
c_{j}'(t) & \equiv\frac{1}{2}[\langle q_{j}(t)q_{j}(0)\rho_{0}\rangle+\langle q_{j}(0)q_{j}(t)\rho_{0}\rangle],\\
c_{j}''(t) & =-\frac{i}{2}[\langle q_{j}(t)q_{j}(0)\rho_{0}\rangle-\langle q_{j}(0)q_{j}(t)\rho_{0}\rangle].
\end{align}
By using Eq.~(\ref{eq:symmetry}), we have
\begin{align}
c_{j}'(t) & =c_{j}'(-t),\\
c_{j}''(t) & =-c_{j}''(-t).
\end{align}
Furthermore, we can define the Fourier transform of the real and imaginary parts of the correlation function of the primary coordinates as
\begin{align}
\tilde{c}_{j}'(\omega) & \equiv\int_{-\infty}^{\infty}dt e^{-i\omega t}c_{j}'(t),\label{eq:C'w}\\
\tilde{c}_{j}''(\omega) & \equiv -i\int_{-\infty}^{\infty}dt e^{-i\omega t}c_{j}''(t).\label{eq:C''w}
\end{align}

In order to calculate $q_{j}(t)$, we use the Heisenberg equation
\begin{align}
\frac{d}{dt}q_{j}(t) & =\frac{1}{i\hbar}[q_{j}(t),H],
\end{align}
and obtain the equation of motion for $q_{j}(t)$ as
\begin{align}
\frac{d}{dt}q_{j}(t) & =\frac{p_{j}(t)}{m_{j}}\left(|g\rangle\langle g|+|e\rangle\langle e|\right).\label{eq:qj}
\end{align}
In the same way, the equation of motion for $p_{j}(t)$ is obtained as
\begin{align}
\frac{d}{dt}p_{j}(t) & =-m_{j}\omega_{j}^{2}q_{j}-|e\rangle\langle e|m_{j}\omega_{j}^{2}d_{j}.\label{eq:pj}
\end{align}
By solving Eqs.~(\ref{eq:qj}) and~(\ref{eq:pj}), we can obtain
\begin{align}
q_{j}(t)&=-d_{j}|e\rangle\langle e|+\left[d_{j}|e\rangle\langle e|+q_{j}(0)\right]\cos(\omega_{j}t)\nonumber\\
&\quad+\frac{p_{j}(0)}{m_{j}\omega_{j}}\sin(\omega_{j}t),\\
p_{j}(t)&=-m_{j}\omega_{j}\left[d_{j}|e\rangle\langle e|+q_{j}(0)\right]\sin(\omega_{j}t)\nonumber\\
&\quad+p_{j}(0)\cos(\omega_{j}t).
\end{align}
Furthermore, we rewrite $q_{j}(t)$ in terms of $a_{j}^{\dagger}$ and $a_{j}$ as
\begin{align}
q_{j}(t) & =d_{j}|e\rangle\langle e|\left[\cos(\omega_{j}t)-1\right]+\frac{a_{j}e^{-i\omega_{j}t}+a_{j}^{\dagger}e^{i\omega_{j}t}}{\sqrt{2m_{j}\omega_{j}}}.
\label{eq:pjt}\end{align}

Substituting Eqs.~(\ref{eq:rhoini}) and~(\ref{eq:pjt}) into Eq.~(\ref{eq:Cjt}), we have
\begin{align}
c_j(t) &= \frac{1}{2 m_j \omega_j\, {\rm Tr}[e^{-\beta H_g}]} \Big[
    \left\langle a_j a_j^\dagger 
    e^{-\beta \sum_l \omega_l a_l^\dagger a_l} \right\rangle e^{-i \omega_j t} \notag\\
&\quad + \left\langle a_j^\dagger a_j 
    e^{-\beta \sum_l \omega_l a_l^\dagger a_l} \right\rangle e^{i \omega_j t}
\Big],
\end{align}
where
\begin{align}
&\left\langle a_{j}a_{j}^{\dagger}e^{-\beta\sum_{l}\omega_{l}a_{l}^{\dagger}a_{l}}\right\rangle \nonumber\\
 & =\Bigg[\sum_{n_j=0}^{\infty}(n_j+1)e^{-\beta\omega_{j}n_j}\Bigg]\prod_{k\neq j}\sum_{n_k=0}^{\infty}e^{-\beta\omega_{k}n_k},
\end{align}
\begin{align}
&\left\langle a_{j}^{\dagger}a_{j}e^{-\beta\sum_{l}\omega_{l}a_{l}^{\dagger}a_{l}}\right\rangle\nonumber\\
 & =\Bigg[\sum_{n_j=0}^{\infty}n_je^{-\beta\omega_{j}n_j}\Bigg]\prod_{k\neq j}\sum_{n_k=0}^{\infty}e^{-\beta\omega_{k}n_k},
\end{align}
\begin{align}
{\rm Tr}\left[e^{-\beta\sum_{l}\omega_{l}a_{l}^{\dagger}a_{l}}\right] =\prod_{k}\sum_{n_k=0}^{\infty}\exp(-\beta\omega_{k}n_k).
\end{align}
Therefore, we have
\begin{align}
c_{j}(t) =\frac{1}{2m_{j}\omega_{j}}\!\!\left[\!\left(\!\frac{2\sum_{n=0}^{\infty}ne^{-\beta\omega_{j}n}}{\sum_{n=0}^{\infty}e^{-\beta\omega_{j}n}}\!+\!1\!\right)\!\cos\omega_{j}t-i\sin\omega_{j}t\right].
\end{align}
Since
\begin{align}
\sum_{n=0}^{\infty}ne^{-\beta\omega_{j}n}  &=\frac{e^{-\beta\omega_{j}}}{\left(1-e^{-\beta\omega_{j}}\right)^{2}},\nonumber\\\label{eq:Sn}
\sum_{n=0}^{\infty}e^{-\beta\omega_{j}n}&=\frac{1}{1-e^{-\beta\omega_{j}}},
\end{align}
 $c_{j}(t)$ can be obtained as
\begin{equation}
c_{j}(t)=\frac{1}{2m_{j}\omega_{j}}\left[\coth\left(\frac{\beta\omega_{j}}{2}\right)\cos(\omega_{j}t)-i\sin(\omega_{j}t)\right].
\end{equation}
According to Eq.~(\ref{eq:Ctxi}),
\begin{align}
C(t) =\sum_{j}g_{j}^{2}\left[\coth\left(\frac{\beta\omega_{j}}{2}\right)\cos(\omega_{j}t)-i\sin(\omega_{j}t)\right].
\end{align}
Replacing the summation by the integral, the TCF reads
\begin{align}
C(t) & =\!\!\int_{0}^{\infty}\!\!d\omega g^{2}(\omega)\rho(\omega)\left[\coth\left(\frac{\beta\omega}{2}\right)\cos(\omega t)-i\sin(\omega t)\right],
\end{align}
and
\begin{align}
C'(t) & =\int_{0}^{\infty}d\omega g^{2}(\omega)\rho(\omega)\coth\left(\frac{\beta\omega}{2}\right)\cos(\omega t), \\
C''(t) & =-\int_{0}^{\infty}d\omega g^{2}(\omega)\rho(\omega)\sin(\omega t),
\end{align}
with $\rho(\omega)$ being the density of the state at frequency $\omega$ in the bath.

According to Eqs.~(\ref{eq:C'w}) and~(\ref{eq:C''w}), we have
\begin{align}
\tilde{C}'(\omega) & =\pi g^{2}(\omega)\rho(\omega)\coth(\frac{\beta\omega}{2}),\\
\tilde{C}''(\omega) & =\pi g^{2}(\omega)\rho(\omega).
\end{align}
The TCF reads
\begin{align}
C(t) & =\frac{1}{\pi}\int_{0}^{\infty}d\omega\tilde{C}''(\omega)\left[\coth\left(\frac{\beta\omega}{2}\right)\cos(\omega t)-i\sin(\omega t)\right],\label{eq:Ct-C''}
\end{align}
and
\begin{align}
C'(t) & =\frac{1}{\pi}\int_{0}^{\infty}d\omega\tilde{C}''(\omega)\coth\left(\frac{\beta\omega}{2}\right)\cos(\omega t),\label{eq:C'-C''} \\
C''(t) & =-\frac{1}{\pi}\int_{0}^{\infty}d\omega\tilde{C}''(\omega)\sin(\omega t),\label{eq:C''-C''}
\end{align}
and the line-shape function is
\begin{align}
g(t) & =\int_{0}^{t}d\tau_{2}\int_{0}^{\tau_{2}}d\tau_{1}C(\tau_{1})\nonumber \\
 & =\frac{1}{\pi}\int_{0}^{\infty}d\omega\frac{\tilde{C}''(\omega)}{\omega^{2}}\bigl\{\coth\left(\frac{\beta\omega}{2}\right)\left[1-\cos(\omega t)\right]\nonumber \\
 & \quad +i\left[\sin(\omega t)-\omega t\right]\bigr\}.
 \label{eq:gt}
\end{align}
Note that the spectral density is defined as $\mathcal{J}(\omega)=\tilde{C}''(\omega)/\pi=g^{2}(\omega)\rho(\omega)$.

\section{Relation between CLS and TCF}
\label{app:cls}

In the cases where the TCF is complex, based on the response-function theory, $R_j$ $(j=1,2,3,4)$ are as follows \cite{Mukamel1999}
\begin{align}
R_{1}\left(t ,T ,\tau \right) & =\mu^{4}e^{-i\omega_{eg}\left(\tau +t \right)}e^{g_{1}\left(t ,T ,\tau \right)},\label{eq:R1}\\
R_{2}\left(t ,T ,\tau \right) & =\mu^{4}e^{-i\omega_{eg}\left(t -\tau \right)}e^{g_{2}\left(t ,T ,\tau \right)},\label{eq:R2}\\
R_{3}\left(t ,T ,\tau \right) & =\mu^{4}e^{-i\omega_{eg}\left(t -\tau \right)}e^{g_{3}\left(t ,T ,\tau \right)},\label{eq:R3}\\
R_{4}\left(t ,T ,\tau \right) & =\mu^{4}e^{-i\omega_{eg}\left(\tau +t \right)}e^{g_{4}\left(t ,T ,\tau \right)},\label{eq:R4}
\end{align}
where
\begin{align}
g_{1}\left(t ,T ,\tau \right) =&-g\left(\tau \right)-g^{*}\left(T \right)-g^{*}\left(t \right)+g\left(\tau +T \right)\nonumber\\
&+g^{*}\left(T +t \right)-g\left(\tau +T +t \right),\label{eq:g1}\\
g_{2}\left(t ,T ,\tau \right) =&-g^{*}\left(\tau \right)+g\left(T \right)-g^{*}\left(t \right)-g^{*}\left(\tau +T \right)\nonumber\\
&-g\left(T +t \right)+g^{*}\left(\tau +T +t \right),\label{eq:g2}\\
g_{3}\left(t ,T ,\tau \right) =&-g^{*}\left(\tau \right)+g^{*}\left(T \right)-g\left(t \right)-g^{*}\left(\tau +T \right)\nonumber\\
&-g^{*}\left(T +t \right)+g^{*}\left(\tau +T +t \right),\label{eq:g3}\\
g_{4}\left(t ,T ,\tau \right) =&-g\left(\tau \right)-g\left(T \right)-g\left(t \right)+g\left(\tau +T \right)\nonumber\\
&+g\left(T +t \right)-g\left(\tau +T +t \right)\label{eq:g4},
\end{align}
and the line-shape function is given by Eq.~(\ref{eq:gt}).

Under the short-time approximation, to the second order of $\tau$ and $t$, i.e., $\cos\omega t\approx1-\omega^{2}t^{2}/2$, $\sin\omega t\approx\omega t$,  
substituting Eqs.~(\ref{eq:Ct-C''})-(\ref{eq:C''-C''}) into Eqs.~(\ref{eq:g1})-(\ref{eq:g4}) yields
\begin{align}
g_{1}\left(t ,T ,\tau \right)
=&-\left(\frac{\tau ^{2}}{2}+\frac{t ^{2}}{2}\right)C'\left(0\right)-\tau t C\left(T \right)-t ^{2}C''\left(T \right)\nonumber\\
&-2t \int_{0}^{T }C''\left(t\right)dt,\\
g_{2}\left(t ,T ,\tau \right)
 =&-\left(\frac{\tau ^{2}}{2}+\frac{t ^{2}}{2}\right)C'\left(0\right)+\tau t C^{*}\left(T \right)-t ^{2}C''\left(T \right)\nonumber\\
 &-2t \int_{0}^{T }C''\left(t\right)dt,\\
g_{3}\left(t ,T ,\tau \right)
=&-\left(\frac{\tau ^{2}}{2}+\frac{t ^{2}}{2}\right)C'\left(0\right)+\tau t C^{*}\left(T \right),\\
g_{4}\left(t ,T ,\tau \right)
 =&-\left(\frac{\tau ^{2}}{2}+\frac{t ^{2}}{2}\right)C'\left(0\right)-\tau t C\left(T \right).
\end{align}
Assuming that the imaginary part of the TCF is equal to zero, e.g. in the high-temperature limit, the response functions in the time domain become
\begin{align}
R_{1}\left(t ,T ,\tau \right) & =R_{4}\left(t ,T ,\tau \right)\nonumber\\
&=\mu^{4}e^{-i\omega_{eg}\left(\tau +t \right)}e^{-\frac{\tau ^{2}+t ^{2}}{2}C\left(0\right)-\tau t C\left(T \right)},\\
R_{2}\left(t ,T ,\tau \right) & =R_{3}\left(t ,T ,\tau \right)\nonumber\\
&=\mu^{4}e^{-i\omega_{eg}\left(t -\tau \right)}e^{-\frac{\tau ^{2}+t ^{2}}{2}C\left(0\right)+\tau t C\left(T \right)}.
\end{align}
Performing the Fourier transforms with respect to $\tau$ and $t$ respectively, i.e.,
\begin{align}
\tilde{R}_{k}\left(\omega_{t},T ,\omega_{\tau}\right)  =&\int d\tau\int dt  e^{i (\omega_{\tau} \tau +\omega_{t}t )} R_{k}\left(t ,T ,\tau \right),\nonumber\\&(k=1,4),\\
\tilde{R}_{l}\left(\omega_{t},T ,\omega_{\tau}\right) =&\int d\tau\int dt  e^{i (-\omega_{\tau} \tau +\omega_{t}t )} R_{l}\left(t ,T ,\tau \right),\nonumber\\&(l=2,3),
\end{align}
the signals in the frequency domain are obtained as
\begin{widetext}
\begin{align}
\tilde{R}_{k}(\omega_{t},T ,\omega_{\tau})=&\frac{2\pi\mu^{4}}{\sqrt{C(0)^{2}-C(T)^{2}}} \exp\left\{\!\!-\frac{C(0)[(\omega_{\tau}-\omega_{eg})^{2}+(\omega_{t}-\omega_{eg})^{2}]-2C(T )(\omega_{\tau}-\omega_{eg})(\omega_{t}-\omega_{eg})}{2[C(0)^{2}-C(T)^{2}]}\right\},(k=1,4),\\
\tilde{R}_{l}(\omega_{t},T ,\omega_{\tau})=&\frac{2\pi\mu^{4}}{\sqrt{C(0)^{2}-C(T)^{2}}} \exp\left\{\!\!-\frac{C(0)[(\omega_{\tau}+\omega_{eg})^{2}+(\omega_{t}-\omega_{eg})^{2}]+2C(T )(\omega_{\tau}+\omega_{eg})(\omega_{t}-\omega_{eg})}{2[C(0)^{2}-C(T)^{2}]}\right\},(l=2,3).
\end{align}
\end{widetext}
Notice that $\tilde{R}_{j}\left(\omega_{t},T ,\omega_{\tau}\right)~(j=1,2,3,4)$ are real as long as the TCF is real. By setting the partial derivatives of $R_{j}\left(\omega_{t},T ,\omega_{\tau}\right)~(j=1,2,3,4)$ with respect to $\omega_{\tau}$ and $\omega_{t}$ equal to zero, respectively, we can obtain the CLS as
\begin{align}
\frac{\partial {\tilde{R}_{k}}}{\partial\omega_{\tau}}=0\Rightarrow\textrm{ CLS}^{R_k}_{\omega_{t}}=\frac{C\left(T \right)}{C\left(0\right)},(k=1,4),\\
\frac{\partial {\tilde{R}_{k}}}{\partial\omega_{t}}=0\Rightarrow \textrm{ CLS}^{R_k}_{\omega_{\tau}}=\frac{C\left(T \right)}{C\left(0\right)},(k=1,4),\\
\frac{\partial {\tilde{R}_{l}}}{\partial\omega_{\tau}}=0\Rightarrow \textrm{ CLS}^{R_l}_{\omega_{t}}=\frac{C\left(T \right)}{C\left(0\right)},(l=2,3),\\
\frac{\partial {\tilde{R}_{l}}}{\partial\omega_{t}}=0\Rightarrow \textrm{ CLS}^{R_l}_{\omega_{\tau}}=\frac{C\left(T \right)}{C\left(0\right)},(l=2,3).
\end{align}
It can be observed that, based on the response-function theory and the short-time approximation, when $C(t)$ is real, all of the CLSs of $\tilde{R}_j~(j=1,2,3,4)$ are equal to the normalized $C(t)$.

\section{Rephasing and Non-Rephasing 2DS}
\label{app:ReNr}

{During the intervals between the two successive probe pulses, the sample interacts with the three control electromagnetic fields, as shown in Fig.~\ref{fig:system}(a). Since in the Schr\"{o}dinger picture, the interaction Hamiltonian between the control fields and the molecule is time-dependent, we transform it to the interaction picture in order to make it time-independent.

In the Schr\"{o}dinger picture, the total Hamiltonian without the system-bath interaction is given by
\begin{equation}
H^{\alpha}=H_{0}^{\alpha}+V^{\alpha},
\end{equation}
where
\begin{equation}
H_{0}^{\alpha}=\sum_{j=1}^{3}\hbar\omega_{j}|e_{j}^{\alpha}\rangle\langle e_{j}^{\alpha}|
\end{equation}
is the Hamiltonian of the four-level system with the energy of the ground state assumed to be zero. Assuming the amplitudes of the control fields are constant, the interaction Hamiltonian, based on the electronic dipole approximation, is given by
\begin{align}
V^{\alpha} & =-\sum_{i>j}\vec{d}_{ij}^{\alpha}\cdot\vec{E}_{ij}(t)\nonumber\\
 & =-\sum_{i>j}\vec{\mu}_{ij}^{\alpha}\cdot\vec{E}_{ij}\left(|e_{i}^{\alpha}\rangle\langle e_{j}^{\alpha}|+|e_{j}^{\alpha}\rangle\langle e_{i}^{\alpha}|\right)\cos\left(\omega_{ij}t\right).
\end{align}
Transforming to the interaction picture with respect to the unitary operater $U_{0}^{\alpha}\left(t\right)=\exp\left(-iH_{0}^{\alpha}t/\hbar\right)$, the efficient Hamiltonian reads
\begin{align}
H_{I}^{\alpha} & =U_{0}^{\alpha\dagger}\left(t\right)V^{\alpha}U^{\alpha}_{0}\left(t\right)\nonumber\\
 & =-\sum_{i>j}\vec{\mu}_{ij}^{\alpha}\cdot\vec{E}_{ij}\left[e^{i\omega_{ij}t}|e_{i}^{\alpha}\rangle\langle e_{j}^{\alpha}|+\textrm{h.c.}\right]\cos\left(\omega_{ij}t\right).\label{eq:H_I_1}
\end{align}
As the control fields are resonant with the corresponding transitions, the efficient Hamiltonian $H_{I}^{\alpha}$ reads
\begin{align}
H_{I}^{\alpha} =-\sum_{i>j}\frac{\vec{\mu}_{ij}^{\alpha}\cdot\vec{E}_{ij}}{2}&[|e_{i}^{\alpha}\rangle\langle e_{j}^{\alpha}|e^{2i\omega_{ij}t}+|e_{i}^{\alpha}\rangle\langle e_{j}^{\alpha}|+\textrm{h.c.}].
\end{align}
According to the rotating-wave approximation, we obtain
\begin{equation}
H_{I}^{\alpha}=-\sum_{i>j}\Omega_{ij}^{\alpha}\left(|e_{i}^{\alpha}\rangle\langle e_{j}^{\alpha}|+|e_{j}^{\alpha}\rangle\langle e_{i}^{\alpha}|\right),
\end{equation}
where the Rabi frequency is defined as $\Omega_{ij}^{\alpha}=-\vec{\mu}_{ij}^{\alpha}\cdot\vec{E}_{ij}/2$. Therefore, the effictive Hamiltonian in the interaction picture is
\begin{equation}
H_{I}^{\alpha}=\Omega_{21}^{\alpha}|e_{2}^{\alpha}\rangle\langle e_{1}^{\alpha}|+\Omega_{31}^{\alpha}|e_{3}^{\alpha}\rangle\langle e_{1}^{\alpha}|+\Omega_{32}^{\alpha}|e_{3}^{\alpha}\rangle\langle e_{2}^{\alpha}|+{\rm h.c.}
\end{equation}}

 Assume that the probe pulses are so strong that the interactions between the three control fields and the system become negligible when the probe pulses are applied. Therefore, in the interaction picture, the Hamiltonian describing the system during a pulse is written as
\begin{equation}
V_{p}^\alpha\left(t\right)=\Omega_{p}\left(t\right)e^{i\vec{k}_{p}\cdot\vec{r}}|e_{1}^\alpha\rangle\langle g^{\alpha}|+{\rm h.c.},
\end{equation}
where the wave vectors of the probe pulses are $\vec{k}_{p}~(p=a,b,c)$, $\vec{r}$ is the spatial location of the molecule, $\Omega_{p}\left(t\right)$ is the Rabi frequency corresponding to the transition $|g \rangle \leftrightarrow |e_1^\alpha \rangle$. Under the square-pulse approximation, the Hamiltonian of the system during a probe pulse becomes
\begin{equation}
V_{p}^\alpha=\Omega_{p}e^{i\vec{k}_{p}\cdot\vec{r}}|e_{1}^\alpha\rangle\langle g^{\alpha}|+{\rm h.c.}
 \end{equation}

The evolution operator under the influence of the pulse can be simplified as
\begin{align}
U_p^\alpha(\delta t_p)
=& e^{-iV_p^\alpha\delta t_p}\nonumber \\
=& \sum_{n=0}^\infty \! \frac{(-i\Omega_p\delta t_p)^n}{n!}
   \!\Big(\!e^{i\vec{k}_p\!\cdot\!\vec{r}}|e_1^\alpha\rangle\langle g^\alpha|
   \!+\! e^{-i\vec{k}_p\!\cdot\!\vec{r}}|g^\alpha\rangle\langle e_1^\alpha|\Big)^{\!n} \nonumber\\
=& -i\sin(\Omega_p\delta t_p)
   \!\left(e^{i\vec{k}_p\!\cdot\!\vec{r}}|e_1^\alpha\rangle\langle g^\alpha|
   + e^{-i\vec{k}_p\!\cdot\!\vec{r}}|g^\alpha\rangle\langle e_1^\alpha|\right)\nonumber\\
   &+ \cos(\Omega_p\delta t_p)
   \!\left(|e_1^\alpha\rangle\langle e_1^\alpha|
   + |g^\alpha\rangle\langle g^\alpha|\right)\nonumber\\
   &+ |e_2^\alpha\rangle\langle e_2^\alpha|
   + |e_3^\alpha\rangle\langle e_3^\alpha|
.
\end{align}
And thus we have
\begin{equation}\label{eq:Upwhole}
\begin{split}
U_{p}^\alpha\left(\delta t_{p}\right)|g^{\alpha}\rangle & \!=\!\cos\left(\Omega_{p}\delta t_{p}\right)|g^{\alpha}\rangle-i\sin\left(\Omega_{p}\delta t_{p}\right)e^{i\vec{k}_{p}\cdot\vec{r}}|e_{1}^\alpha\rangle,\\
U_{p}^\alpha\left(\delta t_{p}\right)|e_{1}^\alpha\rangle & \!=\!\cos\left(\Omega_{p}\delta t_{p}\right)|e_{1}^\alpha\rangle-i\sin\left(\Omega_{p}\delta t_{p}\right)e^{-i\vec{k}_{p}\cdot\vec{r}}|g^{\alpha}\rangle,\\
U_{p}^\alpha\left(\delta t_{p}\right)|e_{2}^\alpha\rangle & =|e_{2}^\alpha\rangle,\\
U_{p}^\alpha\left(\delta t_{p}\right)|e_{3}^\alpha\rangle & =|e_{3}^\alpha\rangle.
\end{split}
\end{equation}
In other words, the pulse will induce the coherent mixing between the ground state and the first excited state, while leave the system unchanged when it is initially in the other excited states.
Considering the pulses are so short that $\Omega_{p}\delta t_{p}\ll1$, to the first-order approximation, Eq.~(\ref{eq:Upwhole}) can be simplified as
\begin{equation}\label{eq:Upfirstorder}
\begin{split}
U_{p}^{\alpha}\left(\delta t_{p}\right)|g^{\alpha}\rangle & =\mathcal{N}_{p}\left(|g^{\alpha}\rangle+\beta_{p}e^{i\vec{k}_{p}\cdot\vec{r}}|e_{1}^{\alpha}\rangle\right),\\
U_{p}^{\alpha}\left(\delta t_{p}\right)|e_{1}^{\alpha}\rangle & =\mathcal{N}_{p}\left(|e_{1}^{\alpha}\rangle+\beta_{p}e^{-i\vec{k}_{p}\cdot\vec{r}}|g^{\alpha}\rangle\right),\\
U_{p}^{\alpha}\left(\delta t_{p}\right)|e_{2}^{\alpha}\rangle & =|e_{2}^{\alpha}\rangle,\\
U_{p}^{\alpha}\left(\delta t_{p}\right)|e_{3}^{\alpha}\rangle & =|e_{3}^{\alpha}\rangle,
\end{split}
\end{equation}
where $\beta_{p}=-i\Omega_{p}\delta t_{p}$, $\mathcal{N}_{p}=(1+\left|\beta_{p}\right|^{2})^{-1/2}$
is the normalization constant.
%    \begin{figure}
%    \includegraphics[width=8.5cm]{suppfig1}\caption{Double-side Feynman diagrams for (a) stimulated emission process and (b) ground-state bleaching process.\label{fig:Feynmandiagrams}}
%    \end{figure}

Correspondingly, the influence of the time-evolution operators in the quantum simulation $U_\textrm{QS}^{\alpha}\left(\tau,0\right)$, $U_\textrm{QS}^{\alpha}\left(T+\tau,\tau\right)$ and $U_\textrm{QS}^{\alpha}\left(t+T+\tau,T+\tau\right)$ on the states can be listed as follows
\begin{equation}\label{eq:UQS}
\begin{split}
U_\textrm{QS}^{\alpha}\left(\tau,0\right)|e_{l}\rangle & =\sum_{j=1}^{3}a_{jl}|e_{j}^{\alpha}\rangle,\\
U_\textrm{QS}^{\alpha}\left(\tau,0\right)|g^{\alpha}\rangle & =|g^{\alpha}\rangle,\\
U_\textrm{QS}^{\alpha}\left(T+\tau,\tau\right)|e_{l}\rangle & =\sum_{j=1}^{3}b_{jl}|e_{j}^{\alpha}\rangle,\\
U_\textrm{QS}^{\alpha}\left(T+\tau,\tau\right)|g^{\alpha}\rangle & =|g^{\alpha}\rangle,\\
U_\textrm{QS}^{\alpha}\left(t+T+\tau,T+\tau\right)|e_{l}\rangle & =\sum_{j=1}^{3}c_{jl}|e_{j}^{\alpha}\rangle,\\
U_\textrm{QS}^{\alpha}\left(t+T+\tau,T+\tau\right)|g^{\alpha}\rangle & =|g^{\alpha}\rangle.
\end{split}
\end{equation}

 In the interaction picture, the wave function at the final stage is given as
\begin{align}
|\psi_I^{\alpha}(\tau,T,t)\rangle=&U_\textrm{QS}^{\alpha}\left(t+T+\tau,T+\tau\right)U_{cI}^{\alpha}U_\textrm{QS}^{\alpha}\left(T+\tau,\tau\right)\nonumber\\
&U_{bI}^{\alpha}U_\textrm{QS}^{\alpha}\left(\tau,0\right)U_{aI}^{\alpha}|\psi_{0}^{\alpha}\rangle.
\end{align}
Setting $|\psi_{0}^{\alpha}\rangle=|g^{\alpha}\rangle$ and using Eq.~(\ref{eq:UQS}), it can be explicitly given as
\begin{align}
&|\psi_I^{\alpha}(\tau,T,t)\rangle  =\mathcal{N}_{a}\mathcal{N}_{b}\mathcal{N}_{c}|g^{\alpha}\rangle\nonumber\\
 & +\mathcal{N}_{a}\left(\mathcal{N}_{b}-1\right)\left(\mathcal{N}_{c}-1\right)\beta_{a}e^{i\vec{k}_{a}\cdot\vec{r}}a_{11}b_{11}\sum_{j=1}^{3}c_{j1}|e_{j}^{\alpha}\rangle\nonumber\\
 & +\mathcal{N}_{a}\left(\mathcal{N}_{b}-1\right)\beta_{a}e^{i\vec{k}_{a}\cdot\vec{r}}a_{11}\sum_{j=1}^{3}b_{j1}\sum_{j'=1}^{3}c_{j'j}|e_{j'}^{\alpha}\rangle\nonumber\\
 & +\mathcal{N}_{a}\left(\mathcal{N}_{c}-1\right)\beta_{a}e^{i\vec{k}_{a}\cdot\vec{r}}\sum_{j=1}^{3}a_{j1}b_{1j}\sum_{j'=1}^{3}c_{j'1}|e_{j'}^{\alpha}\rangle\nonumber\\
 & +\mathcal{N}_{a}\beta_{a}e^{i\vec{k}_{a}\cdot\vec{r}}\sum_{j,j'=1}^{3}a_{j1}b_{j'j}\sum_{s=1}^{3}c_{sj'}|e_{s}^{\alpha}\rangle\nonumber\\
 & +\mathcal{N}_{a}\mathcal{N}_{b}\left(\mathcal{N}_{c}-1\right)\beta_{b}e^{i\vec{k}_{b}\cdot\vec{r}}b_{11}\sum_{j=1}^{3}c_{j1}|e_{j}^{\alpha}\rangle\nonumber\\
 & +\mathcal{N}_{a}\mathcal{N}_{b}\beta_{b}e^{i\vec{k}_{b}\cdot\vec{r}}\sum_{j=1}^{3}b_{j1}\sum_{j'=1}^{3}c_{j'j}|e_{j'}^{\alpha}\rangle\nonumber\\
 & +\mathcal{N}_{a}\mathcal{N}_{b}\mathcal{N}_{c}\beta_{c}e^{i\vec{k}_{c}\cdot\vec{r}}\sum_{j=1}^{3}c_{j1}|e_{j}^{\alpha}\rangle\nonumber\\
 & +\mathcal{N}_{a}\mathcal{N}_{b}\mathcal{N}_{c}\beta_{a}\beta_{b}a_{11}e^{i\left(\vec{k}_{a}-\vec{k}_{b}\right)\cdot\vec{r}}|g^{\alpha}\rangle\nonumber\\
 & +\mathcal{N}_{a}\left(\mathcal{N}_{b}-1\right)\mathcal{N}_{c}\beta_{a}\beta_{c}e^{i\left(\vec{k}_{a}-\vec{k}_{c}\right)\cdot\vec{r}}a_{11}b_{11}|g^{\alpha}\rangle\nonumber\\
 & +\mathcal{N}_{a}\mathcal{N}_{c}\beta_{a}\beta_{c}e^{i\left(\vec{k}_{a}-\vec{k}_{c}\right)\cdot\vec{r}}\sum_{j=1}^{3}a_{j1}b_{1j}|g^{\alpha}\rangle\nonumber\\
 & +\mathcal{N}_{a}\mathcal{N}_{b}\mathcal{N}_{c}\beta_{c}\beta_{b}e^{i\left(\vec{k}_{b}-\vec{k}_{c}\right)\cdot\vec{r}}b_{11}|g^{\alpha}\rangle\nonumber\\
 & +\mathcal{N}_{a}\mathcal{N}_{b}\mathcal{N}_{c}\beta_{a}\beta_{b}\beta_{c}a_{11}e^{i\left(\vec{k}_{a}-\vec{k}_{b}+\vec{k}_{c}\right)\cdot\vec{r}}\sum_{j=1}^{3}c_{j1}|e_{j}^{\alpha}\rangle.
\label{eq:finalstate}
\end{align}

The signal around the transition frequency $\omega_1$ is proportional to $\vec{P}^{\alpha}=\rho_{10}^{\alpha}(\tau,T,t)\vec{\mu}_{01}^{\alpha}+\textrm{c.c.}$, with $\rho_{10}^{\alpha}(\tau,T,t)=\langle e_{1}^{\alpha}|\psi_I^{\alpha}(\tau,T,t)\rangle \langle \psi_I^{\alpha}(\tau,T,t)|g^{\alpha}\rangle$, where

%\begin{align*}
%\langle\psi_{0}|e_{1}\rangle & =\mathcal{N}_{a}\left(\mathcal{N}_{b}-1\right)\left(\mathcal{N}_{c}-1\right)\beta_{a}^{*}e^{-i\vec{k}_{a}\cdot\vec{r}}a_{1,1}^{*}b_{1,1}^{*}c_{1,1}^{*}\\
%& +\mathcal{N}_{a}\left(\mathcal{N}_{b}-1\right)\beta_{a}^{*}e^{-i\vec{k}_{a}\cdot\vec{r}}a_{1,1}^{*}\sum_{j}b_{j,1}^{*}c_{1,j}^{*}\\
%& +\mathcal{N}_{a}\left(\mathcal{N}_{c}-1\right)\beta_{a}^{*}e^{-i\vec{k}_{a}\cdot\vec{r}}\sum_{j}a_{j,1}^{*}b_{1,j}^{*}c_{1,1}^{*}\\
%& +\mathcal{N}_{a}\beta_{a}^{*}e^{-i\vec{k}_{a}\cdot\vec{r}}\sum_{j,j'}a_{j,1}^{*}b_{j',j}^{*}c_{1,j'}^{*}\\
% & +\mathcal{N}_{a}\mathcal{N}_{b}\left(\mathcal{N}_{c}-1\right)\beta_{b}^{*}e^{-i\vec{k}_{b}\cdot\vec{r}}b_{1,1}^{*}c_{1,1}^{*}\\
% & +\mathcal{N}_{a}\mathcal{N}_{b}\beta_{b}^{*}e^{-i\vec{k}_{b}\cdot\vec{r}}\sum_{j}b_{j,1}^{*}c_{1,j}^{*}\\
% & +\mathcal{N}_{a}\mathcal{N}_{b}\mathcal{N}_{c}\beta_{c}^{*}e^{-i\vec{k}_{c}\cdot\vec{r}}c_{1,1}^{*}\\
% & +\mathcal{N}_{a}\mathcal{N}_{b}\mathcal{N}_{c}\beta_{a}^{*}\beta_{b}^{*}\beta_{c}^{*}a_{1,1}^{*}e^{i\left(-\vec{k}_{a}+\vec{k}_{b}-\vec{k}_{c}\right)\cdot\vec{r}}c_{1,1}^{*}
%\end{align*}
\begin{align}
&\langle e_{1}^{\alpha}|\psi_I^{\alpha}(\tau,T,t)\rangle  =\mathcal{N}_{a}\left(\mathcal{N}_{b}-1\right)\left(\mathcal{N}_{c}-1\right)\beta_{a}e^{i\vec{k}_{a}\cdot\vec{r}}a_{11}b_{11}c_{11}\nonumber\\
& +\mathcal{N}_{a}\left(\mathcal{N}_{b}-1\right)\beta_{a}e^{i\vec{k}_{a}\cdot\vec{r}}a_{11}\sum_{j}b_{j1}c_{1j}\nonumber\\
& +\mathcal{N}_{a}\left(\mathcal{N}_{c}-1\right)\beta_{a}e^{i\vec{k}_{a}\cdot\vec{r}}\sum_{j}a_{j1}b_{1j}c_{11}\nonumber\\
& +\mathcal{N}_{a}\beta_{a}e^{i\vec{k}_{a}\cdot\vec{r}}\sum_{j,j'}a_{j1}b_{j'j}c_{1j'}\nonumber\\
 & +\mathcal{N}_{a}\mathcal{N}_{b}\left(\mathcal{N}_{c}-1\right)\beta_{b}e^{i\vec{k}_{b}\cdot\vec{r}}b_{11}c_{11}\nonumber\\
 & +\mathcal{N}_{a}\mathcal{N}_{b}\beta_{b}e^{i\vec{k}_{b}\cdot\vec{r}}\sum_{j}b_{j1}c_{1j}\nonumber\\
 & +\mathcal{N}_{a}\mathcal{N}_{b}\mathcal{N}_{c}\beta_{c}e^{i\vec{k}_{c}\cdot\vec{r}}c_{11}\nonumber\\
 & +\mathcal{N}_{a}\mathcal{N}_{b}\mathcal{N}_{c}\beta_{a}\beta_{b}\beta_{c}a_{11}e^{i\left(\vec{k}_{a}-\vec{k}_{b}+\vec{k}_{c}\right)\cdot\vec{r}}c_{11},
\label{eq:e1psi}
\end{align}
%\begin{align*}
%\langle\psi_{0}|g^{\alpha}\rangle & =\mathcal{N}_{a}\mathcal{N}_{b}\mathcal{N}_{c}\\
% & +\mathcal{N}_{a}\mathcal{N}_{b}\mathcal{N}_{c}\beta_{a}^{*}\beta_{b}^{*}a_{1,1}^{*}e^{i\left(-\vec{k}_{a}+\vec{k}_{b}\right)\cdot\vec{r}}\\
% & +\mathcal{N}_{a}\left(\mathcal{N}_{b}-1\right)\mathcal{N}_{c}\beta_{a}^{*}\beta_{c}^{*}e^{i\left(-\vec{k}_{a}+\vec{k}_{c}\right)\cdot\vec{r}}a_{1,1}^{*}b_{1,1}^{*}\\
% & +\mathcal{N}_{a}\mathcal{N}_{c}\beta_{a}^{*}\beta_{c}^{*}e^{i\left(-\vec{k}_{a}+\vec{k}_{c}\right)\cdot\vec{r}}\sum_{j}a_{j,1}^{*}b_{1,j}^{*}\\
% & +\mathcal{N}_{a}\mathcal{N}_{b}\mathcal{N}_{c}\beta_{c}^{*}\beta_{b}^{*}e^{i\left(-\vec{k}_{b}+\vec{k}_{c}\right)\cdot\vec{r}}b_{1,1}^{*}
%\end{align*}
\begin{align}
&\langle g^{\alpha}|\psi_I^{\alpha}(\tau,T,t)\rangle  =\mathcal{N}_{a}\mathcal{N}_{b}\mathcal{N}_{c}\nonumber\\
 & +\mathcal{N}_{a}\mathcal{N}_{b}\mathcal{N}_{c}\beta_{a}\beta_{b}a_{11}e^{i\left(\vec{k}_{a}-\vec{k}_{b}\right)\cdot\vec{r}}\nonumber\\
 & +\mathcal{N}_{a}\left(\mathcal{N}_{b}-1\right)\mathcal{N}_{c}\beta_{a}\beta_{c}e^{i\left(\vec{k}_{a}-\vec{k}_{c}\right)\cdot\vec{r}}a_{11}b_{11}\nonumber\\
 & +\mathcal{N}_{a}\mathcal{N}_{c}\beta_{a}\beta_{c}e^{i\left(\vec{k}_{a}-\vec{k}_{c}\right)\cdot\vec{r}}\sum_{j}a_{j1}b_{1j}\nonumber\\
 & +\mathcal{N}_{a}\mathcal{N}_{b}\mathcal{N}_{c}\beta_{c}\beta_{b}e^{i\left(\vec{k}_{b}-\vec{k}_{c}\right)\cdot\vec{r}}b_{11}\label{eq:gpsi}.
\end{align}
Thus, the rephasing signal in the direction $-\vec{k}_{a}+\vec{k}_{b}+\vec{k}_{c}$ reads
\begin{align}
\rho_{10}^{\alpha}   =&\langle e_{1}^{\alpha}|\psi_I^{\alpha}(\tau,T,t)\rangle\langle\psi_I^{\alpha}(\tau,T,t)|g^{\alpha}\rangle\nonumber\\
  =&e^{i\left(-\vec{k}_{a}+\vec{k}_{b}+\vec{k}_{c}\right)\cdot\vec{r}}\Big[\mathcal{N}_{a}^{2}\mathcal{N}_{b}^{2}\mathcal{N}_{c}^{2}\beta_{a}^{*}\beta_{b}^{*}\beta_{c}a_{11}^{*}c_{11}\nonumber\\
 & +\mathcal{N}_{a}^{2}\mathcal{N}_{b}\mathcal{N}_{c}\left(\mathcal{N}_{b}-1\right)\left(\mathcal{N}_{c}-1\right)\beta_{a}^{*}\beta_{b}\beta_{c}^{*}a_{11}^{*}b_{11}^{*}b_{11}c_{11}\nonumber\\
 & +\mathcal{N}_{a}^{2}\mathcal{N}_{b}\mathcal{N}_{c}\left(\mathcal{N}_{b}-1\right)\beta_{a}^{*}\beta_{b}\beta_{c}^{*}a_{11}^{*}b_{11}^{*}\sum_{j}b_{j1}c_{1j}\nonumber\\
 & +\mathcal{N}_{a}^{2}\mathcal{N}_{b}\mathcal{N}_{c}\left(\mathcal{N}_{c}-1\right)\beta_{a}^{*}\beta_{b}\beta_{c}^{*}\sum_{j}a_{j1}^{*}b_{1j}^{*}b_{11}c_{11}\nonumber\\
 & +\mathcal{N}_{a}^{2}\mathcal{N}_{b}\mathcal{N}_{c}\beta_{a}^{*}\beta_{b}\beta_{c}^{*}\sum_{j}a_{j1}^{*}b_{1j}^{*}\sum_{j'}b_{j'1}c_{1j'}\Big],
%& \simeq e^{i\left(-\vec{k}_{a}+\vec{k}_{b}+\vec{k}_{c}\right)\cdot\vec{r}}\mathcal{N}_{a}^{2}\mathcal{N}_{b}^{2}\mathcal{N}_{c}^{2}\beta_{a}^{*}\beta_{b}^{*}\beta_{c}a_{1,1}^{*}c_{1,1}\\
% & +e^{i\left(-\vec{k}_{a}+\vec{k}_{b}+\vec{k}_{c}\right)\cdot\vec{r}}\mathcal{N}_{a}^{2}\mathcal{N}_{b}\mathcal{N}_{c}\beta_{a}^{*}\beta_{b}\beta_{c}^{*}\sum_{j}a_{j,1}^{*}b_{1,j}^{*}\sum_{j'}b_{j',1}c_{1,j'}\\
\label{eq:rprho10}
\end{align}
while
\begin{equation}\label{eq:rprho01}
\rho_{01}^{\alpha}=\langle g^{\alpha}|\psi_I^{\alpha}(\tau,T,t)\rangle\langle\psi_I^{\alpha}(\tau,T,t)|e_{1}^{\alpha}\rangle=0.
\end{equation}
Notice that in the rephasing direction $\rho_{01}^{\alpha}\neq(\rho_{10}^{\alpha})^\ast$.

The non-rephasing signal in the direction $\vec{k}_{a}-\vec{k}_{b}+\vec{k}_{c}$ reads
\begin{align}
\rho_{10}^{\alpha}  =&\langle e_{1}^{\alpha}|\psi_I^{\alpha}(\tau,T,t)\rangle\langle\psi_I^{\alpha}(\tau,T,t)|g^{\alpha}\rangle\nonumber\\
  =&e^{i\left(\vec{k}_{a}-\vec{k}_{b}+\vec{k}_{c}\right)\cdot\vec{r}}\Big[\mathcal{N}_{a}^{2}\mathcal{N}_{b}^{2}\mathcal{N}_{c}^{2}\beta_{a}\beta_{b}\beta_{c}a_{11}c_{11}\nonumber\\
 & +\mathcal{N}_{a}^{2}\mathcal{N}_{b}\mathcal{N}_{c}\left(\mathcal{N}_{b}-1\right)\left(\mathcal{N}_{c}-1\right)\beta_{a}\beta_{b}^{*}\beta_{c}^{*}b_{11}^{*}a_{11}b_{11}c_{11}\nonumber\\
 & +\mathcal{N}_{a}^{2}\mathcal{N}_{b}\mathcal{N}_{c}\left(\mathcal{N}_{b}-1\right)\beta_{a}\beta_{b}^{*}\beta_{c}^{*}b_{11}^{*}a_{11}\sum_{j}b_{j1}c_{1j}\nonumber\\
 & +\mathcal{N}_{a}^{2}\mathcal{N}_{b}\mathcal{N}_{c}\left(\mathcal{N}_{c}-1\right)\beta_{a}\beta_{b}^{*}\beta_{c}^{*}b_{11}^{*}\sum_{j}a_{j1}b_{1j}c_{11}\nonumber\\
 & +\mathcal{N}_{a}^{2}\mathcal{N}_{b}\mathcal{N}_{c}\beta_{a}\beta_{b}^{*}\beta_{c}^{*}b_{11}^{*}\sum_{jj'}a_{j1}b_{j'j}c_{1j'}\Big],\label{eq:nrrho10}
%& \simeq \mathcal{N}_{a}^{2}\mathcal{N}_{b}^{2}\mathcal{N}_{c}^{2}\beta_{a}\beta_{b}\beta_{c}e^{i\left(\vec{k}_{a}-\vec{k}_{b}+\vec{k}_{c}\right)\cdot\vec{r}}a_{1,1}c_{1,1}\\
% & +\mathcal{N}_{a}^{2}\mathcal{N}_{b}\mathcal{N}_{c}\beta_{a}\beta_{b}^{*}\beta_{c}^{*}e^{i\left(\vec{k}_{a}-\vec{k}_{b}+\vec{k}_{c}\right)\cdot\vec{r}}b_{1,1}^{*}\sum_{j,j'}a_{j,1}b_{j',j}c_{1,j'}
\end{align}
while
\begin{equation}\label{eq:nrrho01}
\rho_{01}^{\alpha} =\langle g^{\alpha}|\psi_I^{\alpha}(\tau,T,t)\rangle\langle\psi_I^{\alpha}(\tau,T,t)|e_{1}^{\alpha}\rangle\\
 =0.
\end{equation}
Notice that in the non-rephasing direction $\rho_{01}^{\alpha}\neq(\rho_{10}^{\alpha})^\ast$.

{Up to the third order, we can obtain the polarization for the rephasing signal as 
\begin{align}
P_{\rm rp}^\alpha(t,T,\tau) =&\mathcal{N}_{a}^{2}\mathcal{N}_{b}^{2}\mathcal{N}_{c}^{2}\beta_{a}^{*}\beta_{b}^{*}\beta_{c}a_{11}^{*}c_{11}\nonumber\\
  &+\mathcal{N}_{a}^{2}\mathcal{N}_{b}\mathcal{N}_{c}\beta_{a}^{*}\beta_{b}\beta_{c}^{*}\sum_{j}a_{j1}^{*}b_{1j}^{*}\sum_{j'}b_{j'1}c_{1j'},
\label{Prp}
\end{align}
where the first term corresponds to the ground-state bleaching and the second term corresponds to the stimulated emission in the double-sided Feynman diagram, as shown in Fig.~\ref{fig:Feynman_2level}(a).
The polarization for the non-rephasing signal is given as 
\begin{align}
P_{\rm nr}^\alpha(t,T,\tau) =&\mathcal{N}_{a}^{2}\mathcal{N}_{b}^{2}\mathcal{N}_{c}^{2}\beta_{a}\beta_{b}\beta_{c}a_{11}c_{11}\nonumber\\
  &+\mathcal{N}_{a}^{2}\mathcal{N}_{b}\mathcal{N}_{c}\beta_{a}\beta_{b}^{*}\beta_{c}^{*}b_{11}^{*}\sum_{j,j'}a_{j1}b_{j'j}c_{1j'},\label{Pnr}
\end{align}
where the first term corresponds to the ground-state bleaching and the second term corresponds to the stimulated emission in the double-sided Feynman diagram, as shown in Fig.~\ref{fig:Feynman_2level}(b).
}

The 2DS is obtained by double Fourier transforms of the time-domain signal with respect to $\tau$ and $t$, i.e.,
\begin{equation}
\begin{split}
\tilde{P}_{\rm rp}^\alpha(\omega_t,T,\omega_\tau) \equiv {\rm Re}\left(\mathcal{F}_{\tau}\left\{\mathcal{F}_{t}[P_{\rm rp}^\alpha(t,T,\tau)]\right\}\right),\\
\tilde{P}_{\rm nr}^\alpha(\omega_t,T,\omega_\tau) \equiv {\rm Re}\left(\mathcal{F}_{\tau}\left\{\mathcal{F}_{t}[P_{\rm nr}^\alpha(t,T,\tau)]\right\}\right).
\end{split}
\end{equation}

\begin{figure*}
\includegraphics[width=16cm]{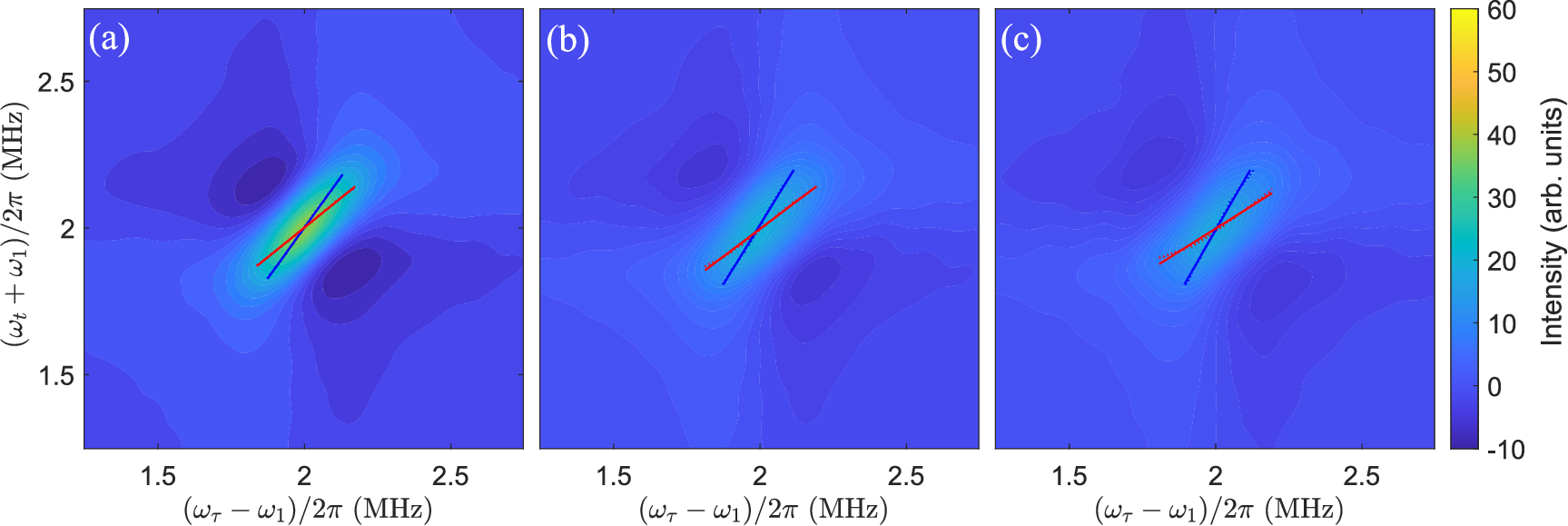}
\caption{The top-right diagonal peak in the rephasing signal of the 2DS simulated by the BET at population times (a) $T=0~\mu$s, (b) $T=4~\mu$s, (c) $T=8~\mu$s. The red (blue) dotted lines are the center lines for $\omega_\tau$ ($\omega_t$). The red (blue) solid lines are the least-squares fitting of the center lines for $\omega_\tau$ ($\omega_t$). The parameters are the same as in Fig.~\ref{fig:rephasing2d}. \label{fig:rephasingfig048}}
\end{figure*}

\begin{figure}
\includegraphics[width=8.5cm]{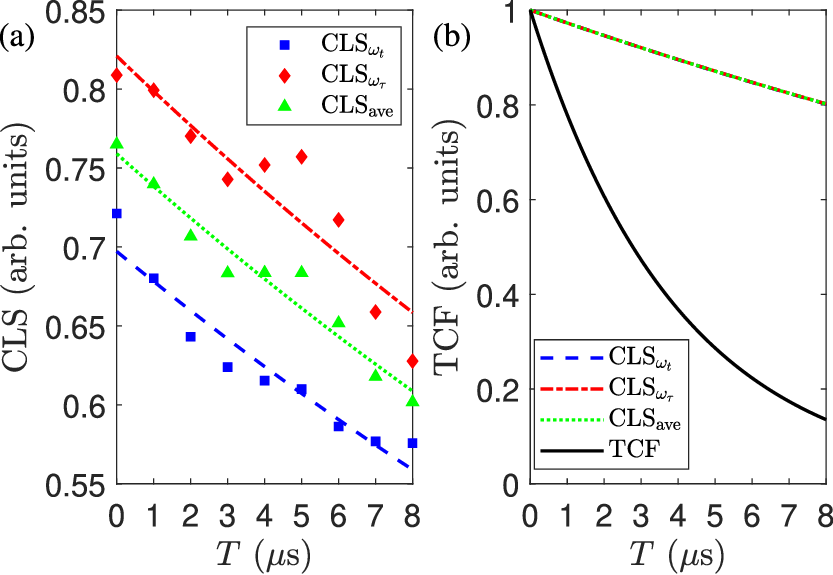}
\caption{(a) The dependence of the CLS, extracted from the rephasing signal of the 2DS, on the population time $T$. The red diamonds (blue squares) are the CLS corresponding to $\omega_\tau$ ($\omega_t$). ${\rm CLS_{ave}}$ is the average of ${\rm CLS}_{\omega_t}$ and ${\rm CLS}_{\omega_\tau}$. The red dashed-dotted line, blue dashed line, and the green dotted line are the corresponding fitted results. (b) The normalized CLSs are compared with the TCF denoted by the black solid line. The parameters are the same as in Fig.~\ref{fig:rephasing2d}. \label{fig:rephasingfig4}}
\end{figure}

We utilize the BET to simulate the 2DS in the rephasing direction as shown in Fig.~\ref{fig:rephasingfig048}. Based on the obtained center line, we can numerically fit the CLS with an exponential function by the least-square fitting as shown in Fig.~\ref{fig:rephasingfig4}(a). We further compare CLS$_{\omega_t}$, CLS$_{\omega_\tau}$, and their average CLS$_{\rm avg}$ with the TCF in Fig.~\ref{fig:rephasingfig4}(b). We may find that although these three CLSs coincide and decay with time, their decay rate is significantly slower than that of the TCF.

\begin{figure*}
\includegraphics[width=16cm]{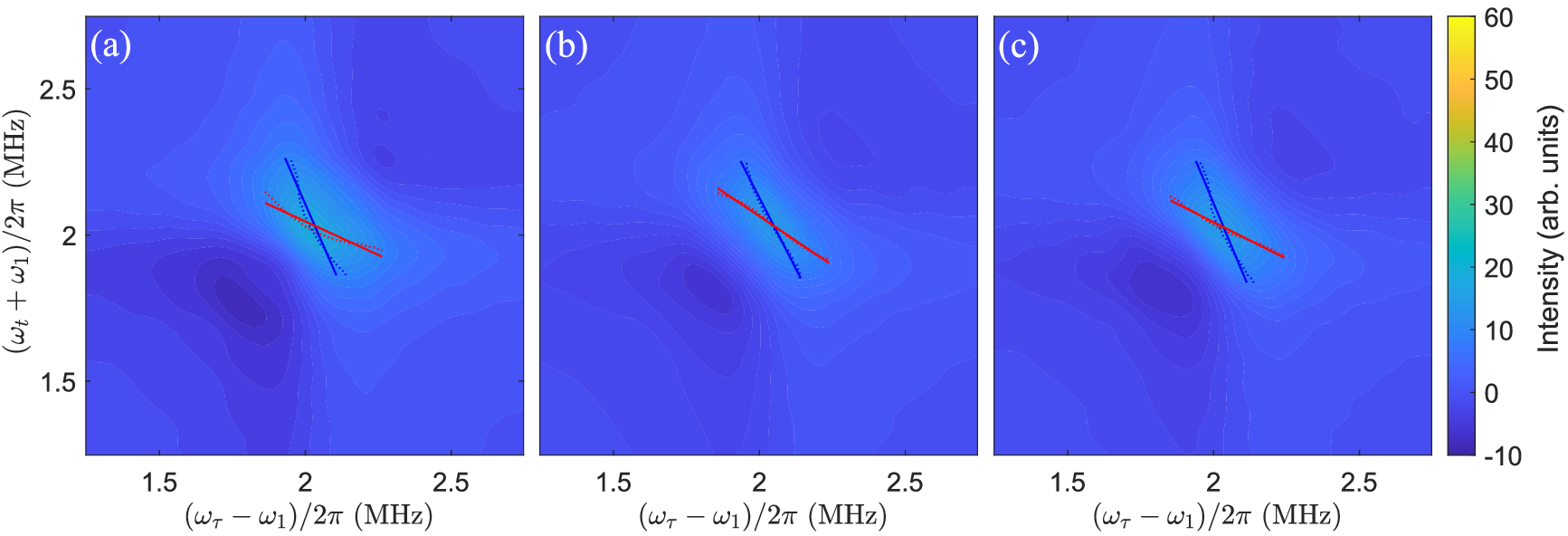}
\caption{The top-right diagonal peak in the non-rephasing signal of the 2DS simulated by the BET at population times  (a) $T=0~\mu$s, (b) $T=4~\mu$s, (c) $T=8~\mu$s. The red (blue) dotted lines are the center lines for $\omega_\tau$ ($\omega_t$). The red (blue) solid lines are the least-squares fitting of the center lines for $\omega_\tau$ ($\omega_t$). The parameters are the same as in Fig.~\ref{fig:rephasing2d}. \label{fig:nonrephasingfig048}}
\end{figure*}

\begin{figure}
\includegraphics[width=8.5cm]{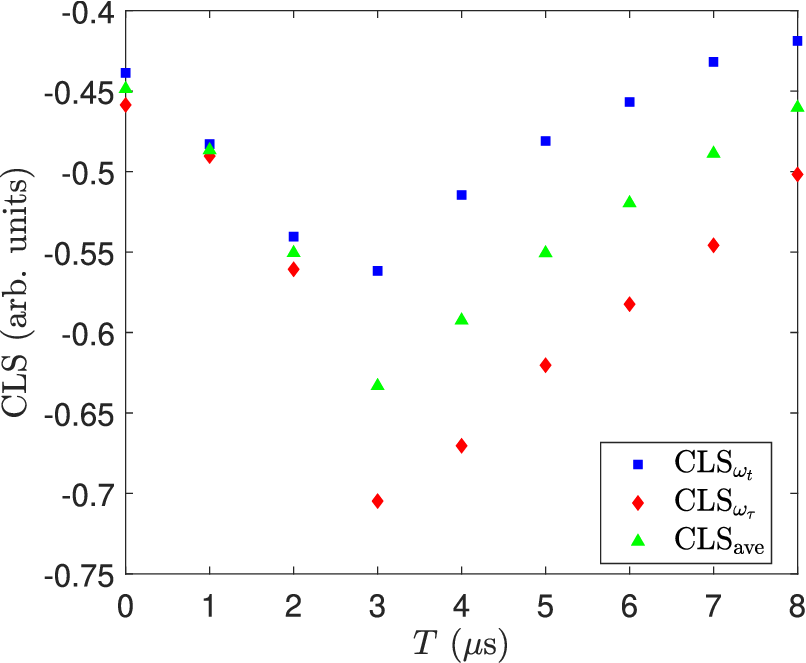}
\caption{The dependence of the CLS, extracted from the non-rephasing signal of the 2DS, on the population time $T$. The red diamonds (blue squares) are the CLS corresponding to $\omega_\tau$ ($\omega_t$). ${\rm CLS_{ave}}$ is the average of ${\rm CLS}_{\omega_t}$ and ${\rm CLS}_{\omega_\tau}$. The parameters are the same as in Fig.~\ref{fig:rephasing2d}. \label{fig:nonrephasingfig4}}
\end{figure}

Then, we utilize the BET to simulate the 2DS in the nonrephasing direction as shown in Fig.~\ref{fig:nonrephasingfig048}. Based on the obtained center line, we can numerically fit the center line by the least-square fitting as shown in Fig.~\ref{fig:nonrephasingfig4}(a).
Surprisingly, the three CLSs are negative. And they all decrease at the first stage while they rise afterwards. Therefore, it is not reasonable to fit the CLS by an exponential function. And thus it may not be utilized to obtain the TCF of the bath.

As shown in Figs.~\ref{fig:rephasingfig048}--\ref{fig:nonrephasingfig4}, the TCF can not be directly obtained from the CLS extracted by the 2DS in either the rephasing or non-rephasing direction. However, as illustrated in Sec.~\ref{sec:CLS}, in the absorptive spectrum obtained by combining the signals from both directions, the TCF extracted using the CLS method agrees quite well with the preset TCF.

\section{Convergence of BET}
\label{app:Convergence}

\begin{figure}
\includegraphics[width=8.5cm]{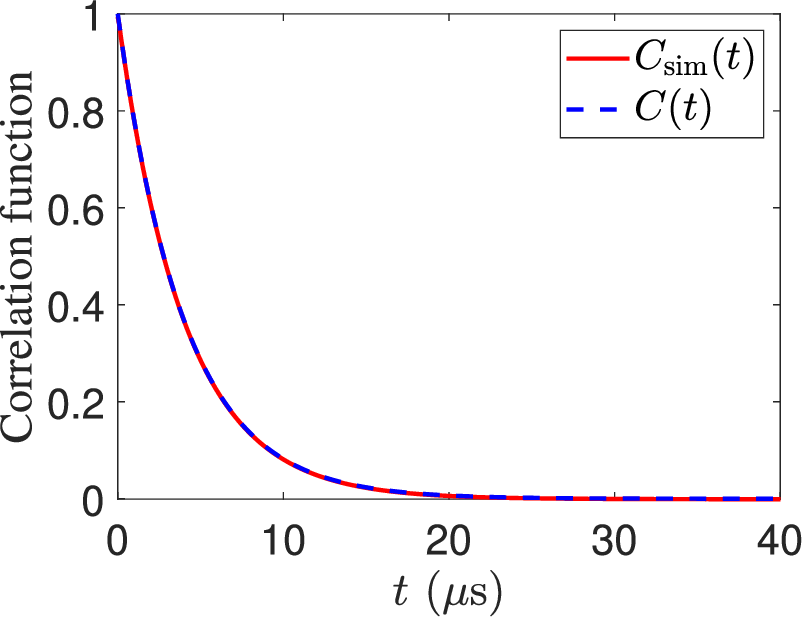}
\caption{Comparison between $C_{\rm sim}(t)$ by the BET and the preset TCF $C(t)$. The parameters are the same as those in Fig.~\ref{fig:rephasing2d}.} \label{fig:Ct_sim}
\end{figure}

In order to simulate the TCF of an exponential form as
\begin{equation}
C(t)=\Delta^2 e^{-|t|/\tau_c},
\end{equation}
we use
\begin{equation}
C_{\rm sim}(t)=\langle B(t)B(0)\rangle=\frac{A^2}{2} \sum_{n=1}^{n_c} \omega_n^2 [F(\omega_n)]^2\cos(\omega t),
\end{equation}
where
\begin{equation}
[F(\omega_n)]^2=\frac{4\omega_0\Delta^2 \tau_c}{\pi\omega_n^2(1+\omega_n^2 \tau_c^2)}.
\end{equation}
The comparison between $C_{\rm sim}(t)$ and the TCF $C(t)$ is shown in Fig.~\ref{fig:Ct_sim}. As we can see, $C_{\rm sim}(t)$ perfectly reproduces $C(t)$.

\begin{figure}
\includegraphics[width=8.5cm]{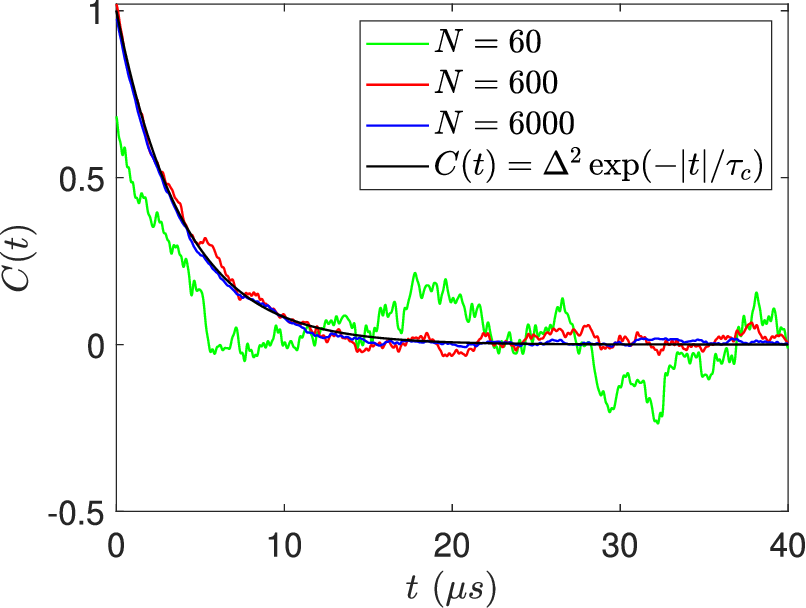}
\caption{Convergence of  $C_{\rm sim}(t)$ with respect to the ensemble size $N$. The green, red, blue solid lines are respectively averaged over an ensemble with $N=60,~600,~6000$ samples. The black solid line is the preset TCF $C(t)$. The other parameters are the same as in Fig.~\ref{fig:rephasing2d}.} \label{fig:Ct_e}
\end{figure}

%Since in the BET, we assume that the average over the ensemble is equivalent to the average over the time, which is justified in the large-ensemble limit. Therefore, it is quite natural to challenge the validity of the BET in finite-ensemble simulation. First of all, in
In Fig.~\ref{fig:Ct_e}, we investigate the convergence of  $C_{\rm sim}(t)$ with respect to the ensemble size $N$. As shown, when the ensemble is small, e.g. $N=60$, $C_{\rm sim}(t)$ fluctuates around the TCF $C(t)$. However, as $N$ increases, e.g. $N=600$, the difference between them becomes smaller. When the ensemble is further increased, e.g. $N=6000$, the difference becomes negligible. 

Furthermore, we investigate the time evolution of the wavefunction as the size of the ensemble increases in Fig.~\ref{fig:stateevolution60_600_6000}. It is shown that as $N$ increases the simulated probability of the wavefunction at the initial state $e_1$ will converge. And there are coherent oscillations at the early stage which are followed by a vanishing steady-state value.
Therefore, we can accurately simulate the open quantum dynamics by a finite-size ensemble when the ensemble is sufficiently large, e.g. $N=6000$.

%\begin{equation}\label{Bt}
%B_j(t)=\alpha \sum_{n=1}^{n_c} F(\omega_n) \omega_n \cos(\omega_n t+\phi_n)
%\end{equation}

%\begin{figure}
%\includegraphics[width=10cm]{betacorrelation_fig2}\caption{} \label{fig:Ct_e}
%\end{figure}

\begin{figure}
\includegraphics[width=8.5cm]{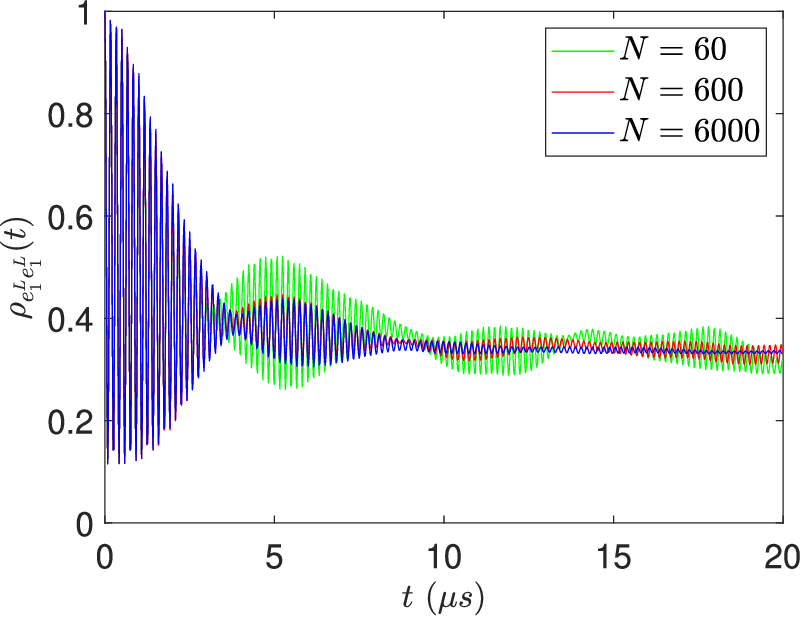}
\caption{Population dynamics of $\vert e_1^{L}\rangle$ with respect to the size of the ensemble $N$. We assume the initial state to be $\vert e_1^{L}\rangle$.  The green, red, blue solid lines are respectively averaged over an ensemble with $N=60,~600,~6000$ samples. The other parameters are the same as in Fig.~\ref{fig:rephasing2d}.} \label{fig:stateevolution60_600_6000}
\end{figure}

In addition, since in our numerical simulation, we simulate the time evolution of the wavefunction by applying the time-evolution operators which are divided into small time steps. To choose the appropriate time step when calculating the chronological integral of the time-evolution operator, we also verify the convergence of the time evolution with respect to the time step $\delta t$ as shown in Fig.~\ref{fig:evolutionsample}. It is shown that as the time step decreases, the results quickly converges and we can hardly discriminate the difference between $\delta t=0.02~\mu$s and $\delta t=0.01~\mu$s. Therefore, the time step can be reasonably chosen as $\delta t=0.02~\mu$s.

\begin{figure}
\includegraphics[width=8.5cm]{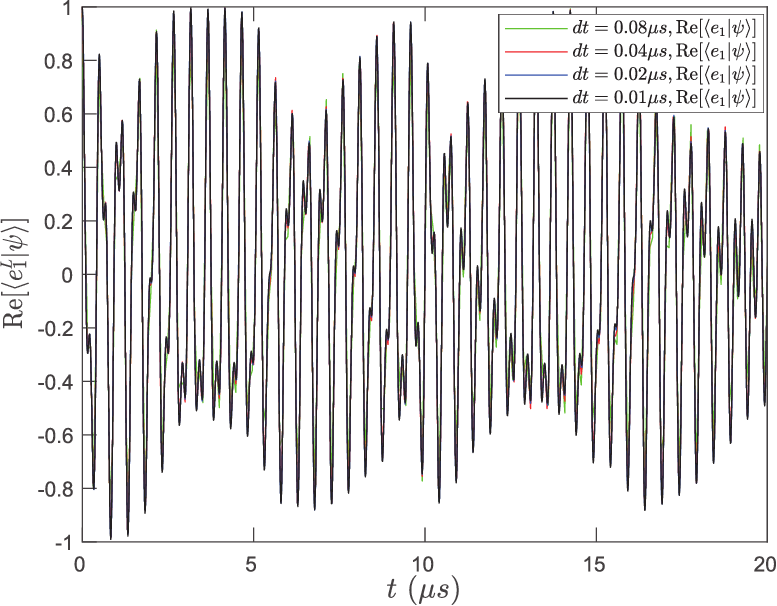}
\caption{The time evolution of $\langle e_1^{L}\vert\psi(t)\rangle$ with respect to the time step $\delta t$. We assume the initial state to be $\vert e_1^{L}\rangle$. The green, red, blue, black solid lines are respectively the time evolutions with $\delta t=0.08,~0.04,~0.02,~0.01~\mu$s. The other parameters are the same as in Fig.~\ref{fig:rephasing2d}.
\label{fig:evolutionsample}}
\end{figure}

\section{Application of BET on Different Platforms with Distinct Environmental Noises}
\label{app:Application}

Experimental platforms exhibit distinct capabilities in engineering system-environment interactions and controlling the spectral characteristics of environmental noise. As long as a given platform can realize a specific type of system-environment coupling, e.g. dephasing noise or amplitude-damping noise, the platform exhibits the potential to simulate a wide range of system-environment spectral densities. 

On the ion-trap platform, it has been theoretically proposed and experimentally demonstrated that both dephasing and amplitude-damping noise can be engineered \cite{Soare2014PRA,Soare2014NP}. 

The dephasing and amplitude-damping noises are constructed via phase and amplitude modulation of the control field, respectively. Both take the form of a summation over discrete frequency components. The dephasing noise associated with the $j$th state is
\begin{equation}\label{eq:IronTrapdephasing}
B_j^{\rm d}(t)=A_j\sum_{n=1}^{n_c}\omega_nF(\omega_n)\cos(\omega_nt+\phi_n^{(j)}),
\end{equation}
while the amplitude-damping noise between states $i$ and $j$ is
\begin{equation}
B_{ij}^{\rm a}(t)=A_{ij}\sum_{n=1}^{n_c}F(\omega_n)\cos(\omega_nt+\phi_n^{(ij)}),
\end{equation}
where $A_j$ and $A_{ij}$ are the global scaling factors, $F(\omega)$ characterizes the strength of each frequency component, $\omega_n=n\omega_0$ is frequency component with base frequency $\omega_0$ and cutoff frequency $n_c\omega_0$, and $\phi_n^{(j)}$, $\phi_n^{(ij)}$ are random phases uniformly distributed in $[0,2\pi)$. The function
$F(\omega)$ can be determined from the correspondence between the correlation function of 
$B_j^{\rm d}(t)$ and $B_{ij}^{\rm a}(t)$ and the TCF of the environment. By choosing an appropriate base frequency $\omega_0$ and the cutoff frequency $n_c\omega_0$, the spectral density can be accurately simulated. 

Since the noise is stochastic, it can also be characterized by its power spectral density, defined as the Fourier transform of its correlation function, i.e., $S(\omega)=\int_{-\infty}^{\infty}dt\langle B(t+\tau)B(\tau) \rangle \exp(i\omega t)$, where $B(t)$ refers to $B_j^{\rm d}(t)$ ($B_{ij}^{\rm a}(t)$) for the dephasing (amplitude-damping) noise. The different forms of $F(\omega)$ for both dephasing and amplitude-damping noises have been experimentally realized and can be found in Tab.~\ref{tab:noise_functions}  \cite{Soare2014PRA,Soare2014NP}.

\begin{table}[htbp]
\renewcommand{\arraystretch}{1.4}
\setlength{\tabcolsep}{0.07cm}
\centering
\caption{$F(\omega)$ for different types of noises with distinct types of power spectral densities.}
\label{tab:noise_functions}
\begin{tabular}{|c|c|c|c|c|c|c|c|c|}
\hline
 & \multicolumn{4}{c|}{Dephasing} & \multicolumn{4}{c|}{Amplitude Damping} \\ \cline{2-9}
 & $1/f^2$ & $1/f$ & White & Ohmic & $1/f^2$ & $1/f$ & White & Ohmic \\ \hline
$F(\omega)$ & 
$\omega^{-2}$ & 
$\omega^{-3/2}$ & 
$\omega^{-1}$ & 
$\omega^{-1/2}$ & 
$\omega^{-1}$ & 
$\omega^{-1/2}$ & 
$\omega^{0}$ & 
$\omega^{1/2}$ \\ \hline
\end{tabular}
\end{table}

%They further implemented the BET experimentally by using IQ modulation to simulate different types of noise. Specifically, they simulate dephasing noise with white and $1/f$ PSD, and amplitude-damping noise with white and $1/f^2$ PSD.

On the NMR platform, dephasing noise comes from the inhomogeneous and non-static magnetic field in NMR systems. The corresponding dephasing noise term is the same as Eq.~(\ref{eq:IronTrapdephasing}). The amplitude-damping noise results from the amplitude fluctuation of the control field, and can be written as \cite{Wang2018NPJQI,zhen2016PRA}
\begin{equation}
B_{ij}^{\rm a}(t)=A_{ij}\sum_{n=1}^{n_c}F(\omega_n)\sin(\omega_nt+\phi_n^{(ij)}).
\end{equation}
The BET has been experimentally implemented in an NMR system using a chloroform sample \cite{Wang2018NPJQI, zhen2016PRA}. Reference~\cite{zhen2016PRA} demonstrated the simulation of amplitude-damping noise with the Ohmic power spectral density. Reference~\cite{Wang2018NPJQI} demonstrated a quantum simulation of dephasing noise with a Drude–Lorentz spectral density. Subsequently, Ref.~\cite{zhang2021FOP} theoretically showed that the same platform can simulate dephasing noise with various system-environment spectral densities, including Ohmic, sub-Ohmic, and super-Ohmic types.

\bibliography{ref}% Produces the bibliography via BibTeX.

\end{document}